\documentclass[journal=apchd5,manuscript=article]{achemso}
\usepackage[version=3]{mhchem} 
\usepackage[T1]{fontenc} 
\usepackage{amsmath,amsthm,amssymb,amsfonts}
\usepackage[english]{babel}
\usepackage{mathtools}
\usepackage{graphicx}
\usepackage{upgreek}
\usepackage[separate-uncertainty = true]{siunitx}
    \DeclareSIPrefix\micro{\ensuremath{\upmu}}{-6}
\usepackage{hyperref}
\usepackage{subcaption}
\usepackage[dvipsnames]{xcolor}
\usepackage{soul}
\usepackage{caption}
\usepackage{lineno}

\newcommand{\PL}{PL}

\newcommand{\ML}{1L-WSe$_2$}

\author{Tobias Bucher}
\affiliation{Institute of Solid-State Physics, Friedrich Schiller University Jena, 07743 Jena, Germany}
\alsoaffiliation{Institute of Applied Physics, Friedrich Schiller University Jena, 07745 Jena, Germany}
\alsoaffiliation{Abbe Center of Photonics, Friedrich Schiller University Jena, 07745 Jena, Germany}
\email{tobias.bucher@uni-jena.de}
\author{Jingshi Yan}
\affiliation{ARC Centre of Excellence for Transformative Meta-Optical Systems (TMOS), Research School of Physics, Australian National University, Canberra ACT 2601, Australia}
\author{Jan Sperrhake}
\affiliation{Institute of Applied Physics, Friedrich Schiller University Jena, 07745 Jena, Germany}
\alsoaffiliation{Abbe Center of Photonics, Friedrich Schiller University Jena, 07745 Jena, Germany}
\author{Zlata Fedorova}
\affiliation{Institute of Solid-State Physics, Friedrich Schiller University Jena, 07743 Jena, Germany}
\alsoaffiliation{Institute of Applied Physics, Friedrich Schiller University Jena, 07745 Jena, Germany}
\alsoaffiliation{Abbe Center of Photonics, Friedrich Schiller University Jena, 07745 Jena, Germany}
\author{Mostafa Abasifard}
\affiliation{Institute of Solid-State Physics, Friedrich Schiller University Jena, 07743 Jena, Germany}
\alsoaffiliation{Institute of Applied Physics, Friedrich Schiller University Jena, 07745 Jena, Germany}
\alsoaffiliation{Abbe Center of Photonics, Friedrich Schiller University Jena, 07745 Jena, Germany}
\author{Rajeshkumar Mupparapu}
\affiliation{Institute of Applied Physics, Friedrich Schiller University Jena, 07745 Jena, Germany}
\alsoaffiliation{Abbe Center of Photonics, Friedrich Schiller University Jena, 07745 Jena, Germany}
\author{Haitao Chen}
\affiliation{ARC Centre of Excellence for Transformative Meta-Optical Systems (TMOS), Research School of Physics, Australian National University, Canberra ACT 2601, Australia}
\author{Emad Najafidehaghani}
\affiliation{Institute of Physical Chemistry, Friedrich Schiller University Jena, 07743 Jena, Germany}
\author{Khosro Zangeneh Kamali}
\affiliation{ARC Centre of Excellence for Transformative Meta-Optical Systems (TMOS), Research School of Physics, Australian National University, Canberra ACT 2601, Australia}
\author{Antony George}
\affiliation{Institute of Physical Chemistry, Friedrich Schiller University Jena, 07743 Jena, Germany}
\alsoaffiliation{Abbe Center of Photonics, Friedrich Schiller University Jena, 07745 Jena, Germany}
\author{Mohsen Rahmani}
\affiliation{ARC Centre of Excellence for Transformative Meta-Optical Systems (TMOS), Research School of Physics, Australian National University, Canberra ACT 2601, Australia}
\author{Thomas Pertsch}
\affiliation{Institute of Applied Physics, Friedrich Schiller University Jena, 07745 Jena, Germany}
\alsoaffiliation{Abbe Center of Photonics, Friedrich Schiller University Jena, 07745 Jena, Germany}
\alsoaffiliation{Fraunhofer Institute for Applied Optics and Precision Engineering IOF, 07745 Jena, Germany}
\alsoaffiliation{Max Planck School of Photonics, Germany}
\author{Andrey Turchanin}
\affiliation{Institute of Physical Chemistry, Friedrich Schiller University Jena, 07743 Jena, Germany}
\alsoaffiliation{Abbe Center of Photonics, Friedrich Schiller University Jena, 07745 Jena, Germany}
\alsoaffiliation{Jena Center for Soft Matter (JCSM), 07743 Jena, Germany}
\author{Dragomir N. Neshev}
\affiliation{ARC Centre of Excellence for Transformative Meta-Optical Systems (TMOS), Research School of Physics, Australian National University, Canberra ACT 2600, Australia}
\author{Isabelle Staude}
\affiliation{Institute of Solid-State Physics, Friedrich Schiller University Jena, 07743 Jena, Germany}
\alsoaffiliation{Institute of Applied Physics, Friedrich Schiller University Jena, 07745 Jena, Germany}
\alsoaffiliation{Abbe Center of Photonics, Friedrich Schiller University Jena, 07745 Jena, Germany}

\title{Valley-dependent emission patterns enabled by plasmonic nanoantennas}

\keywords{Valley-momentum coupling, directional emission, monolayer TMDs, valleytronics, plasmonics, nanoantennas}

\begin{document}
\newpage

\begin{abstract}
Selective control over the emission pattern of valley-polarized excitons in monolayer transition metal dichalcogenides is crucial for developing novel valleytronic, quantum information, and optoelectronic devices. While significant progress has been made in directionally routing photoluminescence from these materials, key challenges remain: notably, how to link routing effects to the degree of valley polarization, and how to distinguish genuine valley-dependent routing from spin-momentum coupling - an optical phenomenon related to electromagnetic scattering but not the light source itself. In this study, we address these challenges by experimentally and numerically establishing a direct relationship between the intrinsic valley polarization of the emitters and the farfield emission pattern, enabling an accurate assessment of valley-selective emission routing. We report valley-selective manipulation of the angular emission pattern of monolayer tungsten diselenide mediated by gold nanobar dimer antennas at cryogenic temperature. Experimentally, we study changes in the system's emission pattern for different circular polarization states of the excitation, demonstrating a valley-selective circular dichroism in photoluminescence of $6\%$. These experimental findings are supported by a novel numerical approach based on the principle of reciprocity, which allows modeling valley-selective emission in periodic systems. We further show numerically, that these valley-selective directional effects are a symmetry-protected property of the nanoantenna array owing to its extrinsic chirality for oblique emission angles, and can significantly be enhanced when tailoring the distribution of emitters. This renders our nanoantenna-based system a robust platform for valleytronic processing.
\end{abstract}


\section*{Introduction}
\begin{figure}[t!]
	\centering
	\includegraphics[width=8.255cm]{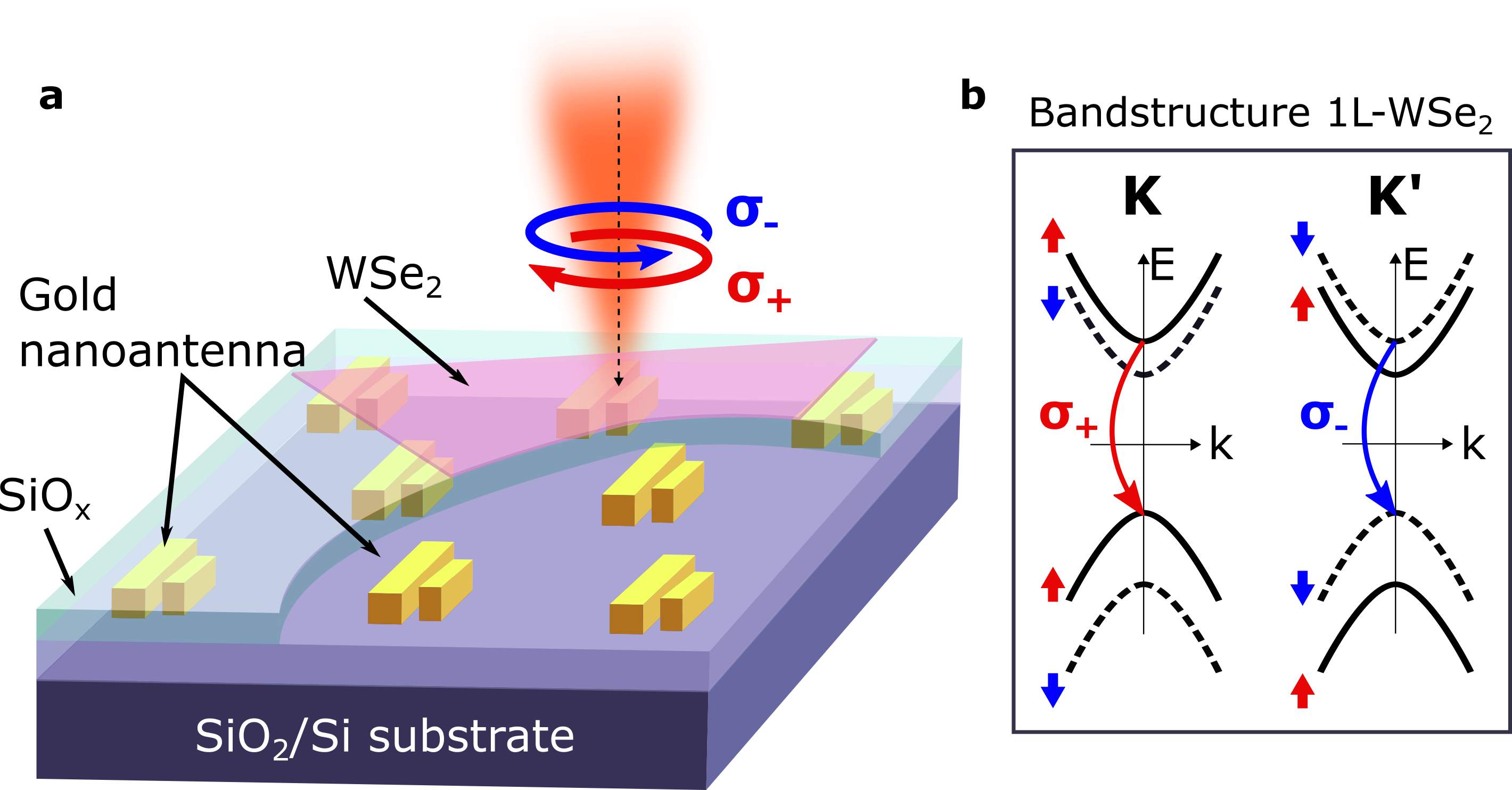}
    {\phantomsubcaption\label{subfig:sketch_a}}
    {\phantomsubcaption\label{subfig:sketch_b}}
	\caption{\textbf{Nanoantenna concept for valley-routing.} (\subref{subfig:sketch_a}) Schematic of a hybrid system consisting of monolayer tungsten disulfide placed on top of an array of gold nanobar dimer antennas. In the monolayer, excitons can be selectively excited in specific valleys using circularly polarized light, as depicted in (\subref{subfig:sketch_b}). Upon radiative decay, valley-excitons act as dipolar nearfield source for the plasmonic nanoantenna, facilitating directional multipolar interference and the corresponding valley-dependent emission directionality.}
	\label{fig:sketch}
\end{figure}

Two-dimensional semiconducting transition metal dichalcogenides (2D-TMDs) possess unique optoelectronic properties, which have propelled them in the spotlight of research in photonics and material science during more than a decade.~\cite{Wang2012electronics, Wang2018colloquium, Schaibley2016valleytronics, Ahmed2017two} Among them are a strong direct-bandgap photoluminescence (PL)~\cite{Mak2010atomically, Splendiani2010emerging} and a high second-order nonlinear susceptibility in the monolayer phase,~\cite{Malard2013observation, Wang2015giant} as well as a pronounced excitonic response at room temperature.~\cite{Mak2012tightly, Ugeda2014giant} Furthermore, the valley pseudospin in 2D-TMDs introduces a new binary degree of freedom for electrons that may be utilized to encode information, paving the way for novel approaches in information processing and storaging.~\cite{Xu2014spin, Mak2014valley, Vitale2018valleytronics, Gong2018nanowire, Liu2019valleytronics, Li2020roomtemperature, Long2022helical, Chen2022chiralitydependent, Shreiner2022electrically, Rong2023spin, Jiang2025chiral} The valley pseudospin arises from multiple energetically degenerate but spin-selective band extrema, the so-called \emph{valleys}, in the conduction and valence bands of a crystal. These valleys form at the direct bandgaps located at the corners of the Brillouin zone, where carriers occupy one of the two subsets (K or K' valleys) depending on their spin state. Additionally, the optical selection rules become valley-dependent, allowing spin-polarized valleys to be selectively addressed and read out using circularly polarized light.~\cite{Mak2012control, Zeng2012valley} The degree of valley polarization (DOVP) reflects the contrast between exciton densities in different valleys and is typically measured through the circular polarization of the emitted PL. While the instantaneous DOVP can reach values close to $\pm 1$, strong intervalley electron-hole exchange interaction leads to a fast valley depolarization, thus limiting the time available for logical processing, transporting and detecting the valley information even at cryogenic temperatures.~\cite{Timmer2024ultrafast, Carvalho2017intervalleyphonon, Yu2014intervalleyeh}\\
Photonic nanostructures offer intriguing opportunities for interfacing 2D-TMDs with light at the nanoscale.~\cite{Krasnok2018review, Mupparapu2020integration, Li2021lightvalleyreview} In particular, they have proven their potential to contribute to solutions for reducing the valley depolarization via various mechanisms. One strategy is to enhance the circular polarization contrast by using chiral nanoparticles,~\cite{Kim2020single} chiral assemblies of metallic nanoparticles,~\cite{Wen2020sandwich} chiral~\cite{Li2018MS-superchiral-nf, Wu2019rtmodulation, Guddala2019metasurfaceqwp} or achiral~\cite{Liu2023controlling} metasurfaces, and other tailored designs.~\cite{Li2021lightvalleyreview} Another approach uses engineered nanostructures to achieve valley-selective directional coupling of valley-polarized excitons or of their emitted light.~\cite{Gong2018nanowire, Yang2019spinvalley, Hu2019coherent, Chervy2018spp, Sun2019groove, Rong2020Rashba, Chen2022chiralitydependent, Huang2024directing}\\
However, most experimentally studied structures for valley routing\,\,\textendash\,\,whether based on propagating surface plasmon polaritons~\cite{Chervy2018spp} and guided modes,~\cite{Gong2018nanowire, Yang2019spinvalley, Chen2022chiralitydependent} or extended modes in metasurfaces~\cite{Sun2019groove, Hu2019coherent} and photonic crystals~\cite{Rong2020Rashba, Wang2020phcslab, Huang2024directing}\,\,\textendash\,\,typically have large footprints of several square microns. The large size of the suggested structures is problematic considering the high integration densities that would be ultimately required for valleytronic devices. A solution to this problem is provided by plasmonic nanoantennas, which are well known for their ability to shape the emission patterns of localized sources.~\cite{Vercruysse2014vantenna, Hancu2014multipolar, Kruk2014spinpolarized, Wen2020sandwich, Zheng2023electron} Plasmonic nanoantennas have also been demonstrated to facilitate free-space emission routing for rotating electric dipole sources, scattering light into different angular directions depending on the rotation direction of the nearfield source.~\cite{Chen2018routing, Wen2020sandwich} Importantly, this approach directly applies to the concept of valley routing, as rotating electric dipoles accurately model the emission from valley-polarized excitons in 2D-TMDs.~\cite{Bucher2024influence, Lamprianidis2022directional, Liu2023controlling, Sun2019groove, Hu2019coherent, Gong2018nanowire}\\
An ideal valley-routing device should scatter PL from valley-polarized excitons into distinct directions based on the dipole's rotation direction, while preserving circular polarization in the farfield to faithfully reflect the underlying DOVP. A natural approach to assess such functionality is through angle- and polarization-resolved measurements of the emitted light. However, interpreting these measurements can be misleading due to the interplay between spin- and valley-dependent effects. Spin-momentum coupling, for instance, can produce directional emission patterns linked to circular polarization regardless of the emitter’s internal valley state.~\cite{Kruk2014spinpolarized, Zheng2023electron} As a result, such routing effects do not reliably indicate the DOVP. Additionally, the scattering process itself can alter the polarization state of the emitted light in a complex manner,~\cite{FernandezCorbaton2013role, ZambranaPuyalto2016tailoring, Bucher2024influence} further complicating interpretation. In this study, we address these challenges by establishing a direct connection between the intrinsic DOVP of the material and the farfield emission characteristics, enabling an accurate assessment of valley-selective emission routing.\\
This connection is established by profound experimental and numerical analysis of valley-selective directional light emission from monolayer tungsten diselenide (\ML{}) placed on top of a plasmonic nanoantenna array as sketched in~\autoref{subfig:sketch_a}.
The nanoantennas consist of two parallel nanobars with different sizes, which have been designed to support electric dipolar and electric quadrupolar resonances, respectively, at the operation wavelength. Chen et al.~\cite{Chen2018routing} numerically demonstrated that the multipolar interference between these resonances leads to emission routing when both nanobars are simultaneously excited by a rotating electric dipole source, with the direction of emission being linked to the rotation direction of the dipole.\\
As a first step, we show that the subwavelength-volume nanoantennas produce distinct angular scattering patterns upon illumination with circularly polarized white light of opposite handedness. We quantify this difference using a normalized contrast metric referred to as angular circular dichroism (CD). Unlike traditional CD measurements involving chiral molecules or structures, the observed effects here are extrinsic, as the nanoantennas themselves are achiral and the asymmetry arises purely from directional scattering.\\
Next, we exploit the valley-dependent selection rules in 1L-WSe$_2$ to generate a pronounced DOVP using circularly polarized excitation as shown in~\autoref{subfig:sketch_b}. We then analyze the directionality of the resulting PL mediated by the nanoantennas, focusing on the angular CD under circularly polarized and unpolarized detection schemes. By comparing both detection modes, we distinguish between purely electromagnetic scattering effects and those that genuinely reflect the underlying DOVP in the material.\\
We further support our experimental analysis with numerical calculations of the angular CD in PL. To accurately calculate the emission of valley-polarized excitons located within the periodic nanoantenna array, we employed a reciprocity-principle-based emission model. Importantly, our calculations show that valley-selective directional scattering is critically linked to the interplay of several nearfield coupling effects within the joint system. On the one hand, by systematically varying the extent of the monolayer in our simulations, we find that valley-selective emitters yield larger magnitudes of the calculated angular CD when positioned in closer proximity to the nanoantenna. On the other hand, we show that the observed directional effects result from extrinsic chirality mediated by the asymmetric nanobar dimer, as no such directional scattering is present in periodic arrays composed of single nanobars.\\
Our numerical analysis reveals how the nearfield polarization around resonant nanoantennas governs the angle- and polarization-dependent farfield PL of nearby valley-selective emitters, providing fundamental insights for the design and optimization of valleytronic devices.
\section*{Nanoantenna fabrication and optical characterization}
\begin{figure}[t!]
    \centering
	\includegraphics[width=15cm]{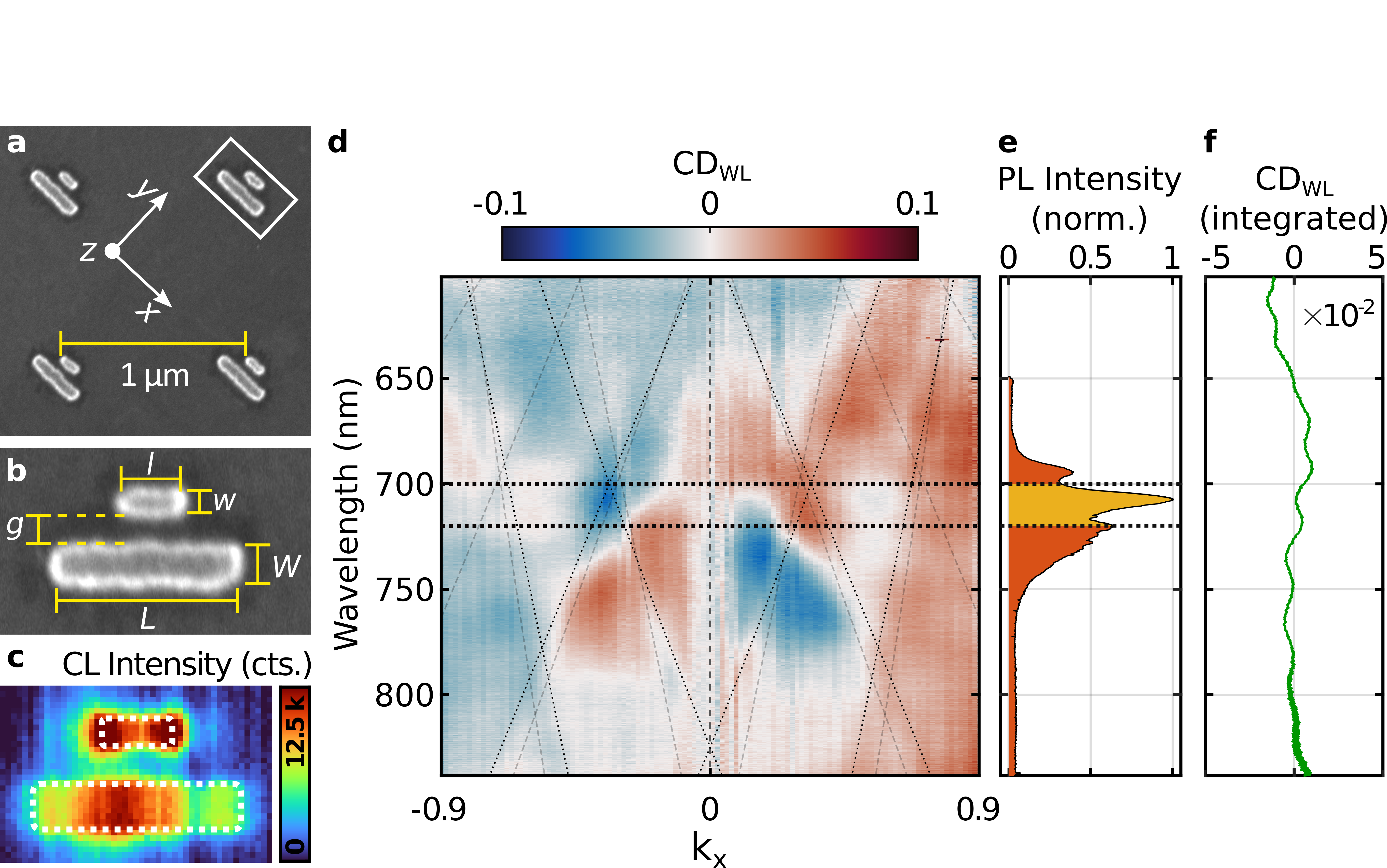}
    {\phantomsubcaption\label{subfig:sample_a}}
    {\phantomsubcaption\label{subfig:sample_b}}
    {\phantomsubcaption\label{subfig:sample_c}}
    {\phantomsubcaption\label{subfig:sample_d}}
    {\phantomsubcaption\label{subfig:sample_e}}
    {\phantomsubcaption\label{subfig:sample_f}}
	\caption{
		\textbf{Optical characterization of the nanoantenna array.} (\subref{subfig:sample_a}) Top-view scanning electron micrograph of a fabricated plasmonic double-bar nanoantenna array. (\subref{subfig:sample_b}) Close-up of a single nanoantenna as indicated by the white box in \subref{subfig:sample_a}) and definition of nanoantenna parameters. (\subref{subfig:sample_c}) CL scan of a nanoantenna from a similar array as shown above. The white dotted lines indicate the nanobar footprints. The CL signal corresponds to a wavelength range of \SIrange{700}{720}{\nano\metre}. (\subref{subfig:sample_d}) Measured angular $\text{CD}_{\text{WL}}$ spectra retrieved from the back-scattered light intensities $I^{\sigma^\pm \text{inc.}}_{\text{WL}}(k_x,\lambda)$ upon illumination with $\sigma^{\pm}$ polarized light. The thin lines indicate the grating orders of the periodic nanoantenna array assuming refractive indices of $n=1.28$ (dashed) and $n=1.65$ (dotted). The thick dotted lines indicate the spectral range of detection of CL imaging.(\subref{subfig:sample_e}) Normalized cryo-PL spectrum of \ML{} on bare substrate. (\subref{subfig:sample_f}) Measured circular dichroism retrieved from the angle-integrated intensities $I^{\sigma^\pm \text{inc.}}_{\text{WL}}(\lambda)$.}
	\label{fig:sample}
\end{figure}

Following the design proposed by Chen et al.,~\cite{Chen2018routing} we fabricated hybrid structures consisting of \ML{} placed on an array of gold nanoantennas. The nanoantennas, each consisting of two parallel nanobars of different sizes, were fabricated on an oxidized silicon wafer (\SI{300}{\nano\metre} oxide layer) using standard electron-beam lithography in combination with gold evaporation and a lift-off process (see Methods for details on the fabrication process). \autoref{subfig:sample_a} shows a scanning electron micrograph of a typical fabricated gold nanoantenna array, where the nanoantennas have a fixed height $H = \SI{40}{\nano\metre}$ and are arranged in a square lattice with a lattice constant $\Lambda = \qty{1}{\micro\metre}$. \autoref{subfig:sample_b} shows a close-up view of the area indicated by the white box.
The small and large nanobar have dimensions of $l\times w = \SI{100}{\nano\metre}\times\SI{40}{\nano\metre}$ and $L\times W = \SI{295}{\nano\metre}\times\SI{65}{\nano\metre}$, respectively, and are separated by a $g = \SI{50}{\nano\metre}$ gap. Additionally, we have fabricated several nanoantenna arrays with varying lengths $l$ and $L$, allowing us to sweep the resonance wavelengths of the short and long nanobars, respectively. To minimize potential changes in the carrier relaxation dynamics within the subsequently transferred \ML{}, which may result from charge-transfer, dipole-dipole interaction, or plasmonic quenching,~\cite{Bender2023spectroscopic, Tugchin2023photoluminescence} we coated the fabricated nanoantenna arrays with a $\SI{15}{\nano\metre}$ thick layer of silicon dioxide using plasma-enhanced chemical vapor deposition.\\
We then investigated the optical nearfield response of the coated nanoantennas using cathodoluminescence (CL) imaging. By exciting the nanostructure with a focused electron-beam, thus generating emitted photons via cascaded processes, this technique provides information on the optical nearfields with super-optical resolution down to few tens of nanometres.~\cite{Knight2012aluminum, Coenen2014directional} In~\autoref{subfig:sample_c}, we show the corresponding CL scan of a single nanoantenna from an array similar to the one depicted in~\autoref{subfig:sample_a}, with the CL signal integrated over a wavelength range from \qtyrange[range-units=single]{700}{720}{\nano\metre}, matching the expected trion emission band in \ML{} at cryogenic temperatures. The CL image reveals a distinct nearfield mode profile for each nanobar. For the small nanobar, we observe a nearfield hotspot at each end of the nanobar. This resembles an electric dipole mode profile, with the dipole moment $p_x$ oriented along the nanobar and centered within it. For the large nanobar, we observe two additional nearfield hotspots near its center, resembling the mode profile of a linear quadrupolar mode, with the quadrupole moment $q_{xx}$ oriented along the nanobar and also centered within it.\\
Exciting both multipolar modes within the nanobars with a rotating dipole leads to directional scattering, with light being predominantly emitted into different halfspaces ($x<0$ and $x>0$) depending on the spin-orientation of the rotating dipole.~\cite{Chen2018routing} In our system, the rotating dipole source is realized by valley-selective excitonic emitters in \ML{} created upon circularly polarized excitation. Note that the nanobars each align along the $x$-direction (see~\autoref{subfig:sample_a}).\\
Next, we investigated the circular-polarization-dependent directional scattering of the fabricated nanoantenna array by angle-resolved white light (WL) spectroscopy. For this, we prepared circularly polarized light using a stabilized tungsten-halogen source combined with a linear polarizer and a superachromatic quarter wave plate. We illuminated the sample and collected the reflected light with a 100x/0.88NA objective. Subsequently, by imaging the back-focal plane of the objective onto the slit of an imaging spectrometer, we measured the angular spectra $I^{\sigma^\pm \text{inc.}}_{\text{WL}}(\sin\theta\cos\varphi,\lambda)$ of the light back-scattered from the sample for $\sigma^{\pm}$ polarized illumination, respectively. Here, $(\theta,\varphi)\in[0,\pi/2]\times[0,\pi]$ are the spherical coordinates describing points on a unit sphere and $\tilde{k}_x =\sin\theta\cos\varphi$ is their corresponding projection along the $x$-direction. A detailed discussion on the setup geometry and the influence of the optical components on the polarization state of the excitation and detection is provided in Sec.~S.1 of the Supporting Information.\\
In~\autoref{subfig:sample_d}, we show the respective angular CD defined as the normalized contrast $\text{CD}_{\text{WL}}=(I^{\sigma^+ \text{inc.}}_{\text{WL}}-I^{\sigma^- \text{inc.}}_{\text{WL}})/(I^{\sigma^+ \text{inc.}}_{\text{WL}}+I^{\sigma^- \text{inc.}}_{\text{WL}})$. Between wavelengths of 700 and \qty{720}{\nm} (see thick dotted lines), we observe a pronounced antisymmetric feature with respect to the two halfspaces $\tilde{k}_x<0$ and $\tilde{k}_x>0$, clearly indicating circular-polarization-dependent directional scattering mediated by the nanoantenna array. In particular, we demonstrate a maximum CD of 6.5\% at a wavelength of \qty{705}{\nm}, matching the cryogenic peak emission wavelength of the PL spectrum measured for \ML{} on the bare substrate at \qty{3.8}{\kelvin}, as shown in~\autoref{subfig:sample_e}. We also observe several additional antisymmetric modes between wavelengths of \qty{700}{\nm} and nearly \qty{800}{\nm}. Since the mode dispersion closely follows that of the lattice modes (highlighted by dashed lines), we attribute these features to the interplay between directional scattering by individual nanoantennas and the diffractive grating orders arising from their periodic arrangement. For details on the dispersion of the grating orders, see Sec.~S.2 of the Supporting Information.\\
It is worth to note that both the single nanoantennas and their periodic arrangement within the array have an achiral geometry such that the total CD of the array is expected to be zero. We have verified this by plotting the measured CD as retrieved from the angle-integrated back-scattered intensities $I^{\sigma^\pm \text{inc.}}_{\text{WL}}(\lambda) = \int_{\text{NA}} I^{\sigma^\pm \text{inc.}}_{\text{WL}}(\tilde{k}_x,\lambda)\operatorname{d}\!\tilde{k}_x$. \autoref{subfig:sample_f} shows the resulting CD, which is close to zero across the entire spectral range of our measurements (note the scale of 0.05). The small oscillating deviations are likely caused by the dispersion of the polarization optics used in our experiments. Note further that in this experiment no polarization control was used in detection and the measured non-zero CD is purely a result of changing the illumination polarization which is a necessary condition to enable valley routing.\\
Finally, we synthesized \ML{} on an oxidized silicon wafer by a chemical vapour deposition process using an Knudsen-type effusion cell as described previously by George et al.~\cite{George2019_cvd-growth}. This scalable process allows for a dense coverage of the growth substrate with single crystalline monolayers, which we subsequently transferred from the growth substrate onto the host substrate with the fabricated nanoantenna array using a poly(methyl methacrylate) assisted wet-transfer scheme~\cite{Turchanin2009-one}.
\section*{Photoluminescence study of the hybrid system}
\begin{figure}[b!]
	\centering
	\includegraphics[width=12cm]{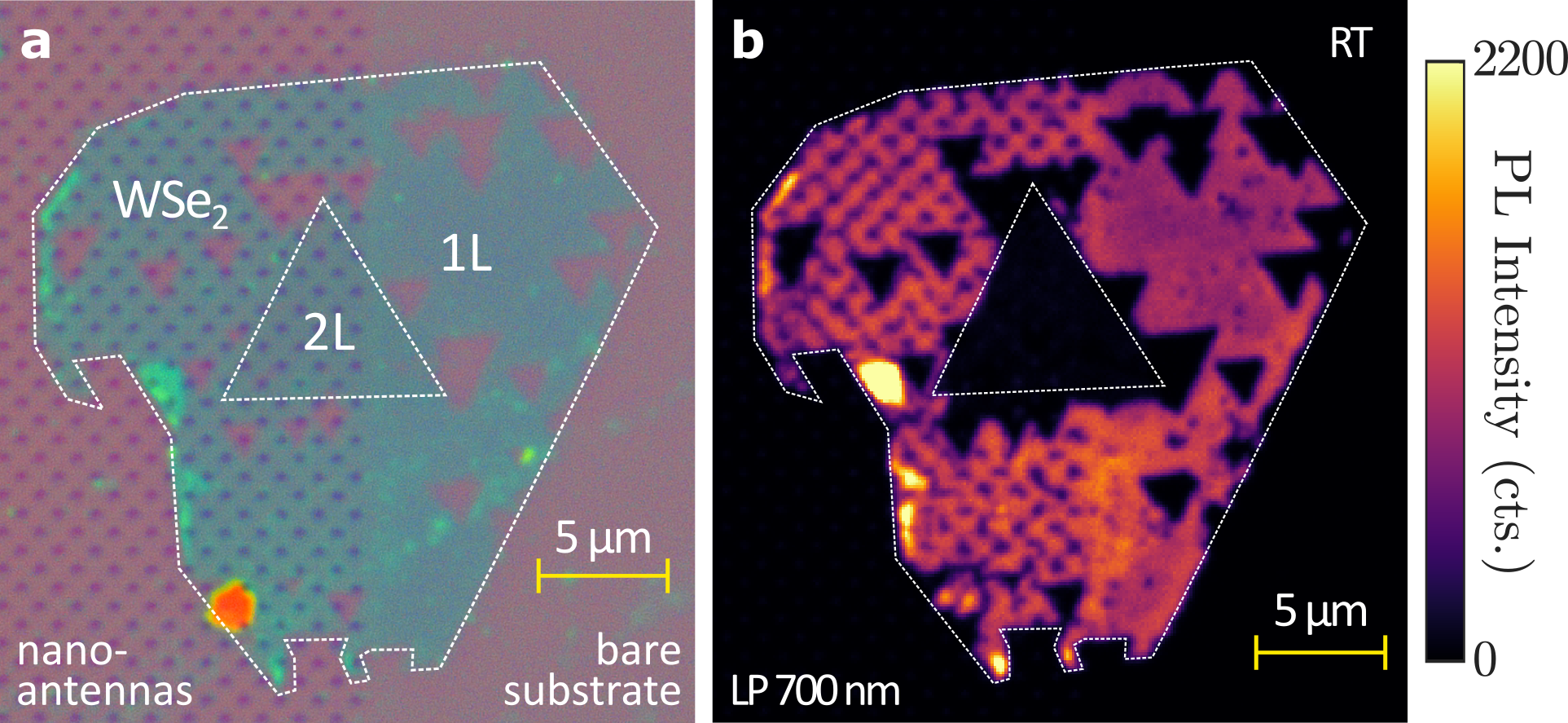}
    {\phantomsubcaption\label{subfig:PL_RT_a}}
    {\phantomsubcaption\label{subfig:PL_RT_b}}
	\caption{\textbf{Room temperature measurements.} (\subref{subfig:PL_RT_a}) Optical microscope image of a \ML{} crystal transferred partly onto a fabricated gold nanoantenna array and partly onto bare substrate. The white dashed lines show the regions with the monolayer (1L) and bilayer (2L) WSe$_2$, as labelled, respectively. (\subref{subfig:PL_RT_b}) Measured confocal scanning microscope image of the PL from the same sample at room temperature using a \SI{700}{\nano\metre} longpass filter in detection. Room temperature PL measurements were conducted without employing circular polarization control of the excitation or detection.}
	\label{fig:PL_RT}
\end{figure}

\begin{figure}[h!]
	\centering
	\includegraphics[width=15.5cm]{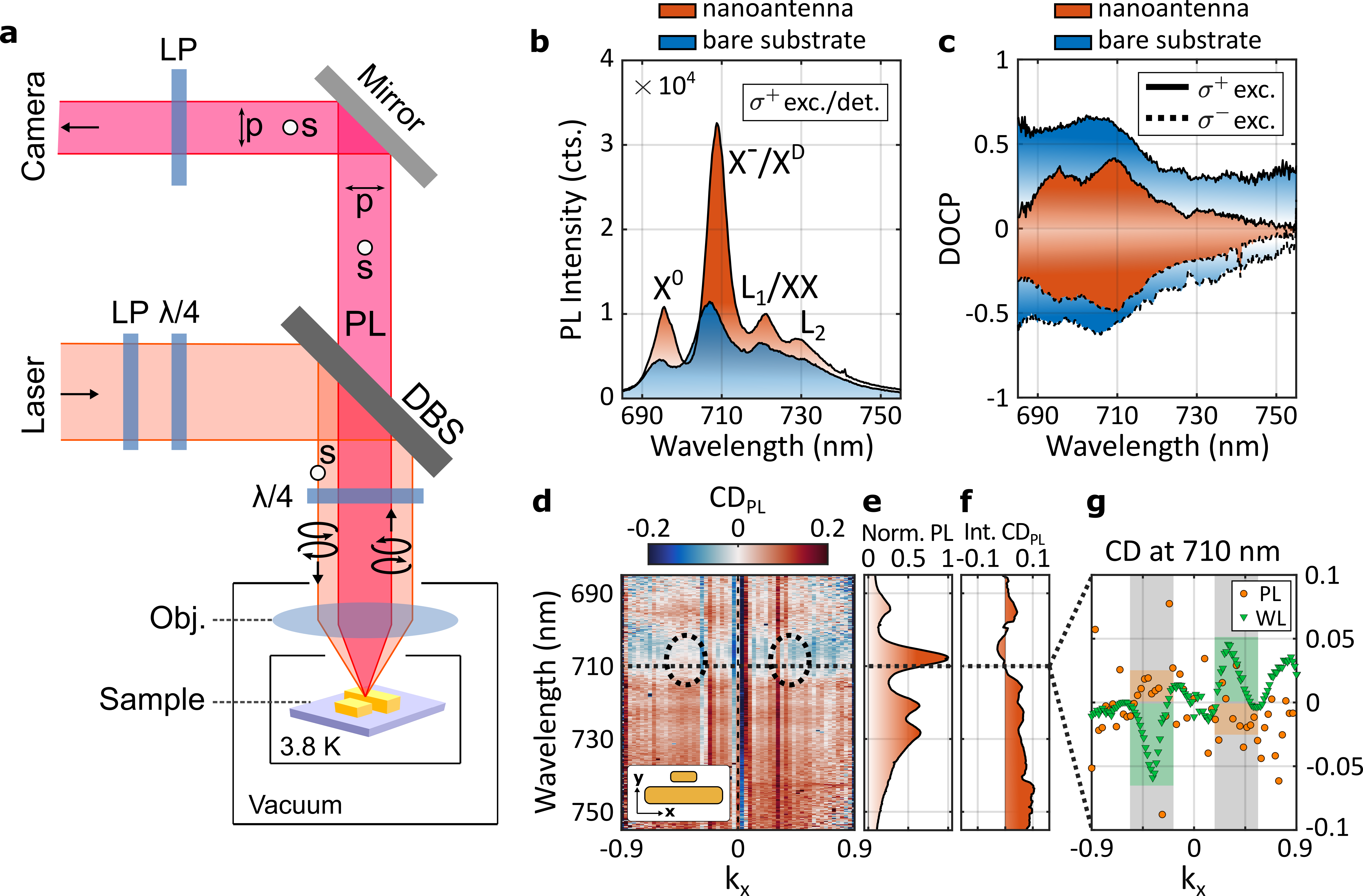}
    {\phantomsubcaption\label{subfig:PL_cryo_a}}
    {\phantomsubcaption\label{subfig:PL_cryo_b}}
    {\phantomsubcaption\label{subfig:PL_cryo_c}}
    {\phantomsubcaption\label{subfig:PL_cryo_d}}
    {\phantomsubcaption\label{subfig:PL_cryo_e}}
    {\phantomsubcaption\label{subfig:PL_cryo_f}}
    {\phantomsubcaption\label{subfig:PL_cryo_g}}
	\caption{\textbf{Cryogenic measurements (T=3.8\,K).} (\subref{subfig:PL_cryo_a}) Sketch of the optical setup for circular-polarization resolved cryogenic PL measurements. (\subref{subfig:PL_cryo_b}) Measured intensity and (\subref{subfig:PL_cryo_c}) DOCP spectra of PL from \ML{} on the gold nanoantenna array (orange curve) and on the bare substrate (blue curve). (\subref{subfig:PL_cryo_d}) Measured CD$_{\text{PL}}$ as retrieved from the momentum-resolved PL intensity spectra $I^{\sigma^\pm \text{inc.}}_{\text{PL}}(k_x,\lambda)$. (\subref{subfig:PL_cryo_e}) Integrated total PL intensity spectrum \mbox{$I_{\text{PL}}(\lambda)$}. (\subref{subfig:PL_cryo_f}) CD$_{\text{PL}}$ as retrieved from the integrated PL intensity spectra $I^{\sigma^\pm \text{inc.}}_{\text{PL}}(\lambda)$. (\subref{subfig:PL_cryo_g}) Cross section of the measured CD$_{\text{PL}}$ (orange dots) and CD$_{\text{WL}}$ (green triangles) at $\SI{710}{\nano\metre}$ wavelength. The dashed ellipses in (\subref{subfig:PL_cryo_d}) and the shaded areas in (\subref{subfig:PL_cryo_g}) were added as a guide to the eye, highlighting the angular range of interest.}
	\label{fig:PLcryo}
\end{figure}

\autoref{subfig:PL_RT_a} shows a true-color optical microscope image of a fabricated nanoantenna array after we transferred a sample of \ML{} to cover parts of the nanoantenna array and of the bare substrate. Note that apart from regions with \ML{}, the crystal includes smaller regions of 2L-WSe\textsubscript{2}, as well as triangular holes (visible as small regions with different contrast). Initially, we investigated the PL emission from the hybrid system at room temperature without employing any polarization control. \autoref{subfig:PL_RT_b} shows a measured confocal PL scan of the same sample area as shown on the left. Further details on the experimental parameters are provided in Methods.
We observe a bright and uniform PL signal from the \ML{} on the bare substrate region associated to the A-excitonic PL. The region of 2L-WSe$_2$ appears darker, due to the drastically reduced PL quantum yield of the indirect bandgap semiconductor. The measured PL spectra of the 1L- and 2L-WSe$_2$ are provided in Sec.~S.3 of the Supporting Information. Note that the PL signal measured from the regions of the triangular holes is indistinguishable from that of the bare substrate. Further, a regular square pattern of small regions with reduced \PL{} intensity coincides with the locations of individual nanoantennas. The reduced PL intensity measured in the farfield is attributed to the interplay of several effects, including the scattering by individual nanoantennas, diffractive grating modes of the array, and Fabry-Pérot modes within the multi-layer substrate. Hence, the interference of multiple contributing modes may lead to a modulation of the farfield PL intensity depending on the geometrical parameters of the sample and the wavelength of emission. For example, we observe enhancement of the farfield PL intensity from \ML{} placed on different nanoantenna arrays with spectrally shifted resonances (see Sec.~S.3 of the Supporting Information).\\
We further investigated the influence of the resonant nanoantennas on the emission decay dynamics of \ML{} using time-resolved PL measurements (see Sec.~S.3 of the Supporting Information). We observe a two-component decay with PL lifetimes of $(2.62\pm 0.05)$\,ns and $(0.37\pm 0.01)$\,ns on the bare substrate, that reduce to $(1.90\pm 0.06)$\,ns and $(0.32\pm 0.01)$\,ns on the nanoantenna array, showing a certain nearfield coupling between the nanoantennas and excitons in \ML{}. Importantly, the nearfield coupling at cryogenic temperatures is expected to be further enhanced, as the exciton energy shifts to higher energies, coinciding with the operational bandwidth of the resonant nanoantenna.\\
Next, we investigated whether the fabricated nanoantenna arrays can directionally route the PL emission from valley-selective excitons in the \ML{}. The induced valley contrast is typically characterized in emission by the degree of circular polarization (DOCP) of PL, defined as $\text{DOCP}=(I_{\text{PL}}^{\sigma^+ \text{det.}}-I_{\text{PL}}^{\sigma^- \text{det.}})/(I_{\text{PL}}^{\sigma^+ \text{det.}}+I_{\text{PL}}^{\sigma^- \text{det.}})$. At room temperature, valley-selective excitation of 1L-WSe$_2$ typically results in a negligible DOCP due to phonon-assisted ultrafast intervalley scattering. Cooling the sample to cryogenic temperatures significantly suppresses the intervalley scattering rate, thereby preserving the circular polarization contrast of the excitation field at the emission level and yielding a pronounced DOCP in PL.\\
Importantly, upon radiative decay, these excitons act as rotating dipolar nearfield sources that drive the nanoantenna at the exciton emission wavelength, with the dipole's rotation sense being determined by the circular polarization of the excitation.~\cite{Bucher2024influence} By design, the nanobar dimer exhibits valley-selective directional farfield interference,~\cite{Chen2018routing} linking the emission direction to the exciton's valley-index and thereby mediating valley-momentum coupling.\\
We performed circular-polarization resolved cryogenic measurements ($T=\qty{3.8}{\kelvin}$) using a commercially available closed-loop liquid helium cryostat (s50, Montana Instruments) equipped with a custom-built back-focal plane imaging setup, as depicted in~\autoref{subfig:PL_cryo_a}.
For excitation, we focused a \SI{633}{\nano\metre} continuous-wave helium-neon laser with an average power of $\SI{100}{\micro\watt}$ on the sample using a 100$\times$/0.88NA objective and collected the PL signal in reflection geometry with the same objective. For polarization control, we used a combination of a linear polarizer (LP) and a quarter-wave plate (QWP) to prepare a linear polarized beam after reflection from the dichroic beam splitter (DBS). Here, the QWP is utilized to compensate the phase shift introduced by the DBS upon reflection. After the DBS, we employed a super-achromatic QWP to prepare a $\sigma^{\pm}$ polarized excitation beam before entering the objective. In detection, the same super-achromatic QWP was used to project the circular-polarized signal into a linear-polarized basis, which is then analyzed by another LP after reflection from a mirror. For a detailed discussion on the polarizing properties of the optical components, see Sec.~S.1 of the Supporting Information.\\
In \autoref{subfig:PL_cryo_b}, we show the experimental PL spectra of \ML{} for $\sigma^+$ polarized excitation and detection measured on the bare substrate (blue curve) and on top of the nanoantenna array (orange curve). In both cases, we observe a multi-peak spectrum that is typical for tungsten-based 1L-TMDs.~\cite{You2015observation, Huang2016probing, Robert2017fine, Courtade2017charged, Molas2019probing, Jindal2025brightened} In ascending wavelength order, the first peak corresponds to the neutral bright exciton ($X^0$), whereas the second peak results from several exciton complexes including negatively charged trions ($X^-$), as well as grey/dark excitons ($D^0$) and trions ($D^-$). Thus, we label this peak as $X^-/D$. Two additional peaks at longer wavelengths correspond to localized defect states, while the third peak also spectrally overlaps with the expected energy of the bi-exciton ($XX$). Thus, we label the third and fourth peak as $L_1/XX$ and $L_2$. A detailed discussion on the excitonic contributions based on multi-Voigt-line fitting is provided in Sec.~S.4 of the Supporting Information.\\
On the nanoantenna array, we observe a higher PL intensity across the entire emission spectrum of \ML{}, likely resulting from excitation and emission enhancement mediated by the nanoantennas (see also Sec.~S.3 of the Supporting Information for time-resolved measurements at room temperature). Additionally, the PL spectrum of \ML{} on the nanoantennas exhibits a redshift by several meV. While such spectral shifts naturally arise from variations across the monolayer sample (compare Fig.~S4b), we find a larger shift for the $X^-/D$ peak, hinting at a potential brightening of dark excitons mediated by the nearfield interaction with the nanoantennas (see Sec.~S.4 of the Supporting Information).\\
Next, we measured the DOCP in PL from \ML{} (see~\autoref{subfig:PL_cryo_c}) on the bare substrate (blue curve) and on the nanoantenna array (orange curve), under $\sigma^+$ (solid curves) and $\sigma^-$ (dashed curves) polarized excitation. In each case, we find the highest DOCP associated with the $X^-/D$ peak, reaching values of 0.65 for the bare substrate and 0.43 for the nanoantenna array. As this reduction in DOCP for \ML{} on the nanoatenna array occurs across the entire emission spectrum, it likely results from scattering by the plasmonic nanoantennas,~\cite{Bucher2024influence} as further analyzed by numerical emission modeling in Sec.~S.9 of the Supporting Information. Note that the slightly asymmetric DOCP in PL on the nanoantenna array is not expected for an achiral structure and may have resulted from small differences in the measurement position between the measurements with different excitation polarization.\\
To quantify the valley-dependence of the emission patterns, we performed angle-resolved spectroscopy of the PL from \ML{} placed on top of the gold nanoantenna array. For these measurements, we used the same detection scheme as in our angle-resolved WL spectroscopy, with the excitation beam widened to cover the same area as for the focused white light illumination (see Methods and Sec.~S.1 of the Supporting Information). \autoref{subfig:PL_cryo_d} shows the corresponding angular CD in PL, defined as $\text{CD}_{\text{PL}}=(I^{\sigma^+ \text{inc.}}_{\text{PL}}-I^{\sigma^- \text{inc.}}_{\text{PL}})/(I^{\sigma^+ \text{inc.}}_{\text{PL}}+I^{\sigma^- \text{inc.}}_{\text{PL}})$, retrieved from the angular PL spectra $I^{\sigma^\pm \text{inc.}}_{\text{PL}}(\tilde{k}_x,\lambda)$. For comparison,~\autoref{subfig:PL_cryo_e} and~\autoref{subfig:PL_cryo_f} show, respectively, the integrated PL intensity spectrum $I_{\text{PL}}(\lambda) = I^{\sigma^+ \text{inc.}}_{\text{PL}}(\lambda) + I^{\sigma^- \text{inc.}}_{\text{PL}}(\lambda)$, where $I^{\sigma^\pm \text{inc.}}_{\text{PL}}(\lambda) = \int_{\text{NA}} I^{\sigma^\pm \text{inc.}}_{\text{PL}}(\tilde{k}_x,\lambda)\operatorname{d}\!\tilde{k}_x$, and the CD$_{\text{PL}}$ calculated from \mbox{$I^{\sigma^\pm \text{inc.}}_{\text{PL}}(\lambda)$}. We find that, in contrast to the case of WL scattering, there is no pronounced directional pattern for the CD$_{\text{PL}}$. Instead, we observe only a minor antisymmetric feature near the DOCP maximum (see dotted line and circles). For clarity, \autoref{subfig:PL_cryo_g} shows the cross-section of the measured CD$_{\text{PL}}$ at a wavelength of \qty{710}{\nm} (orange dots), as well as the cross-section of the CD$_{\text{WL}}$ measured on the nanoantenna array without the monolayer, using white light spectroscopy (green triangles). Compared to the pronounced antisymmetric features observed for CD$_{\text{WL}}$, the distribution for CD$_{\text{PL}}$ is less systematic. Nevertheless, within the angular range discussed previously, as highlighted by the shaded regions, we observe a significant antisymmetric feature in the CD$_{\text{PL}}$, with a magnitude of $2\%$ and its sign opposite to that of the white light result.\\
Note that although the integrated $\mathrm{CD_{PL}}$ must vanish for an achiral nanoantenna array, we observe subtle spectral modulations: the $X^-/X^D$ peak near 709 nm diminishes, while the $X^0$ peak and a broad defect band increase in intensity over successive measurements. These changes are likely caused by slow, photo-induced charge trapping at defect sites and appear to occur on a timescale comparable to the interval between measurements. Importantly, they do not affect the directional $\mathrm{CD_{PL}}$ at any fixed wavelength.\\
\section*{Valley-selective directional scattering}
\begin{figure}[t!]
	\centering
	\includegraphics[width=0.9\textwidth]{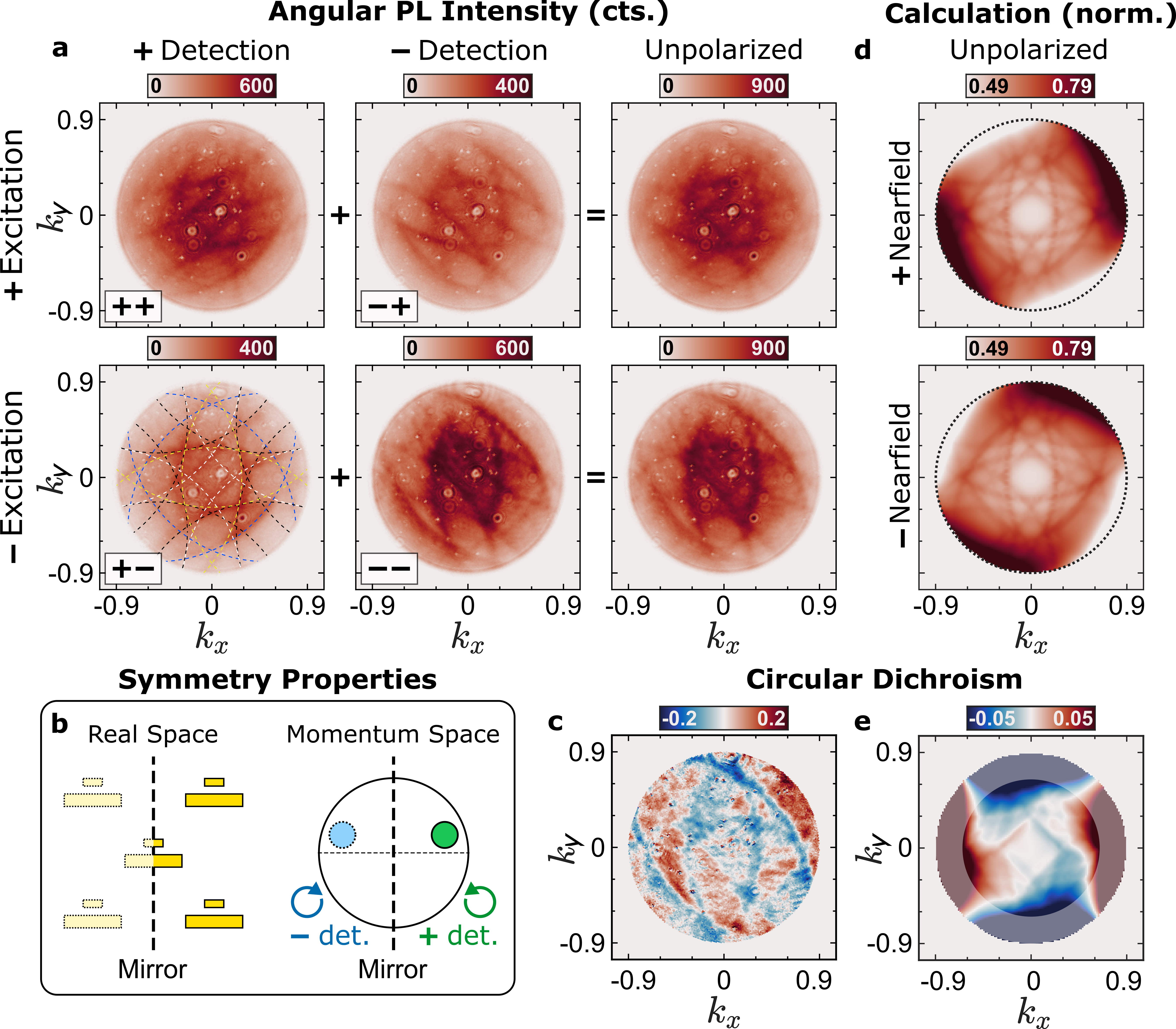}
    {\phantomsubcaption\label{subfig:cryoPL-BFP_a}}
    {\phantomsubcaption\label{subfig:cryoPL-BFP_b}}
    {\phantomsubcaption\label{subfig:cryoPL-BFP_c}}
    {\phantomsubcaption\label{subfig:cryoPL-BFP_d}}
    {\phantomsubcaption\label{subfig:cryoPL-BFP_e}}
	\caption{\textbf{Angular PL intensity distribution and circular dichroism.} (\subref{subfig:cryoPL-BFP_a}) Measured angular PL intensity distribution of \ML{} on the gold nanoantenna array for $\sigma_{+}$ (top row) and $\sigma_{-}$ (bottom row) polarized excitation and $\sigma_{+}$ (left column), $\sigma_{-}$ (middle column), and unpolarized (right column) detection. (\subref{subfig:cryoPL-BFP_b}) Sketch of the system's symmetry properties in real and momentum space. (\subref{subfig:cryoPL-BFP_c}) Measured angular circular dichroism, obtained from the unpolarized case in (\subref{subfig:cryoPL-BFP_a}). (\subref{subfig:cryoPL-BFP_d}) Numerically calculated angular emission intensity distributions obtained from $\sigma_{+}$ (top) and $\sigma_{-}$ (bottom) polarized nearfield intensities, averaged over circular farfield polarizations according to the reciprocity-principle-based approach. (\subref{subfig:cryoPL-BFP_e}) Numerically calculated angular circular dichroism, obtained from (\subref{subfig:cryoPL-BFP_d}).}
	\label{fig:cryoPL-BFP}
\end{figure}

Next, to obtain deeper insight into the observed valley-selective directional asymmetry, we performed polarization-resolved back-focal plane imaging (see Methods for details). \autoref{subfig:cryoPL-BFP_a} shows the measured angular PL intensity, obtained with a $(710\pm 10)$ nm bandpass filter, and for the same measurement location on the nanoantenna array as discussed previously. Specifically, we focused on changes in the angular PL intensity upon switching the circular polarization of the excitation (top and bottom row), and analyzed these emission patterns using a circular-polarized detection basis (left and middle column) and for unpolarized detection (right column).
For simplicity, we introduce a Jones notation to distinguish between these cases as follows: $\sigma^{+}_{\text{det.}}\,|\,\sigma^{-}_{\text{exc.}}\equiv{+-}$, with analogous definitions for all other polarization combinations. These emission patterns show distinct arc patterns resulting from diffractive grating orders that arise due to the periodic nanoantenna arrangement, as discussed in Sec.~S.2 of the Supporting Information. For comparison, we overlaid the calculated diffraction patterns in the bottom left panel in~\autoref{subfig:cryoPL-BFP_a}.\\
In these emission patterns, not all diffractive modes seam to appear with equal intensity, with favored emission directions depending on the circular polarization of detection. For example, for the cases $--$ and $-+$ (i.e. $\sigma^-$ polarized detection), the diffractive modes are pronounced in diagonal $k$-space directions ($k_x\cdot k_y>0$), while for $++$ and $+-$ (i.e. $\sigma^+$ polarized detection) anti-diagonal directions ($k_x\cdot k_y<0$) are favored. This polarization dependence is a signature of spin-momentum coupling mediated by the nanoantenna array. Additionally, these patterns are linked by the system's symmetry in $k$-space, as depicted in~\autoref{subfig:cryoPL-BFP_b}. The emission patterns for opposite detection polarizations appear as mirror images with respect to the $k_y$-axis due to the respective mirror symmetry of the nanoantenna array and the handedness reversal of circularly polarized light under mirror operations. Note that the symmetry of the system is reduced when a valley-specific excitation is introduced. For a given detection polarization, changing the excitation polarization alters the intensity but not the shape of the emission pattern. This intensity difference stems from the finite DOCP in emission from valley-polarized excitons.\\
These observations demonstrate that the angular emission distribution in our system is sensitive to the valley polarization. For unpolarized detection (right column in~\autoref{subfig:cryoPL-BFP_a}), this intensity contrast results in visibly different emission patterns. The respective angular CD distribution in~\autoref{subfig:cryoPL-BFP_c} provides a quantitative measure for these differences, showing a significant contrast across several emission directions. We find angular CD magnitudes up to $15\%$ for specific emission directions, and antisymmetric features with peak values up to $\pm6\%$ along the $k_x$ axis. For a comparison with the results in~\autoref{subfig:PL_cryo_g}, see Sec.~S.5 of the Supporting Information. Importantly, this finite angular CD is a signature of the valley-selectivity in emission with respect to these angular directions. By utilizing valley-selective excitation and unpolarized detection, we demonstrate directional effects arising purely from valley-momentum coupling, distinguishing our results from previous demonstrations of spin-momentum locked emission using emitters without internal (pseudo-)spin.\\
We further performed numerical simulations of the emission from the studied system using a reciprocity-principle-based model.~\cite{Vaskin2019lightemitting,Pal2025dark} This approach bases on the equivalence of placing a single dipolar emitter in the nearfield region of a suitable nanophotonic system and studying its emission behavior in the farfield, with the situation of illuminating the system from the farfield and studying the system's nearfield response. For a detailed description of the numerical model, see Methods. This approach allows us to calculate the radiated power and farfield polarization from incoherent dipolar emitters placed in periodic systems. To model the emission response from 1L-WSe$_2$ on the nanoantenna array, we have calculated the in-plane nearfield distribution on top of the embedding layer upon plane wave illumination. Within this framework, we identify the contribution from a valley-selective emitter in K/K' with the $\sigma^+/\sigma^-$ polarized in-plane nearfield intensity, and associate the polarization of the emitted light with that of the incident plane wave driving the system.\\
By varying the angle of incidence, we obtain the calculated angular emission intensity distributions for unpolarized detection, as depicted in~\autoref{subfig:cryoPL-BFP_d} for $\sigma^+$ polarized (top) and $\sigma^-$ polarized (bottom) nearfield intensities, and the respective CD shown in~\autoref{subfig:cryoPL-BFP_e}. Our calculations show diffractive mode patterns as for our experimental results, giving rise to similar symmetric features as discussed previously. The calculated emission patterns for circular detection polarization are available in Sec.~S.6 of the Supporting Information. The calculated angular CD distribution further confirms the experimentally observed order of magnitude in contrast values for the nanoantenna array homogeneously covered by 1L-WSe$_2$. For small emission angles, we find similar antisymmetric features as previously observed in experiments with an angular CD magnitude of $1\%$ (see also Sec.~S.5 of the Supporting Information). At high emission angles, our model seems to overestimate the emission intensities, and hence contrast values, as such are clearly not observed in experiments. We note that the predictive accuracy of this model is limited for these grazing emission angles, as we did not radiometrically normalize the power radiated into different angles.\\
\section*{Emitter distribution}
\begin{figure}[t!]
	\centering
	\includegraphics[width=0.9\textwidth]{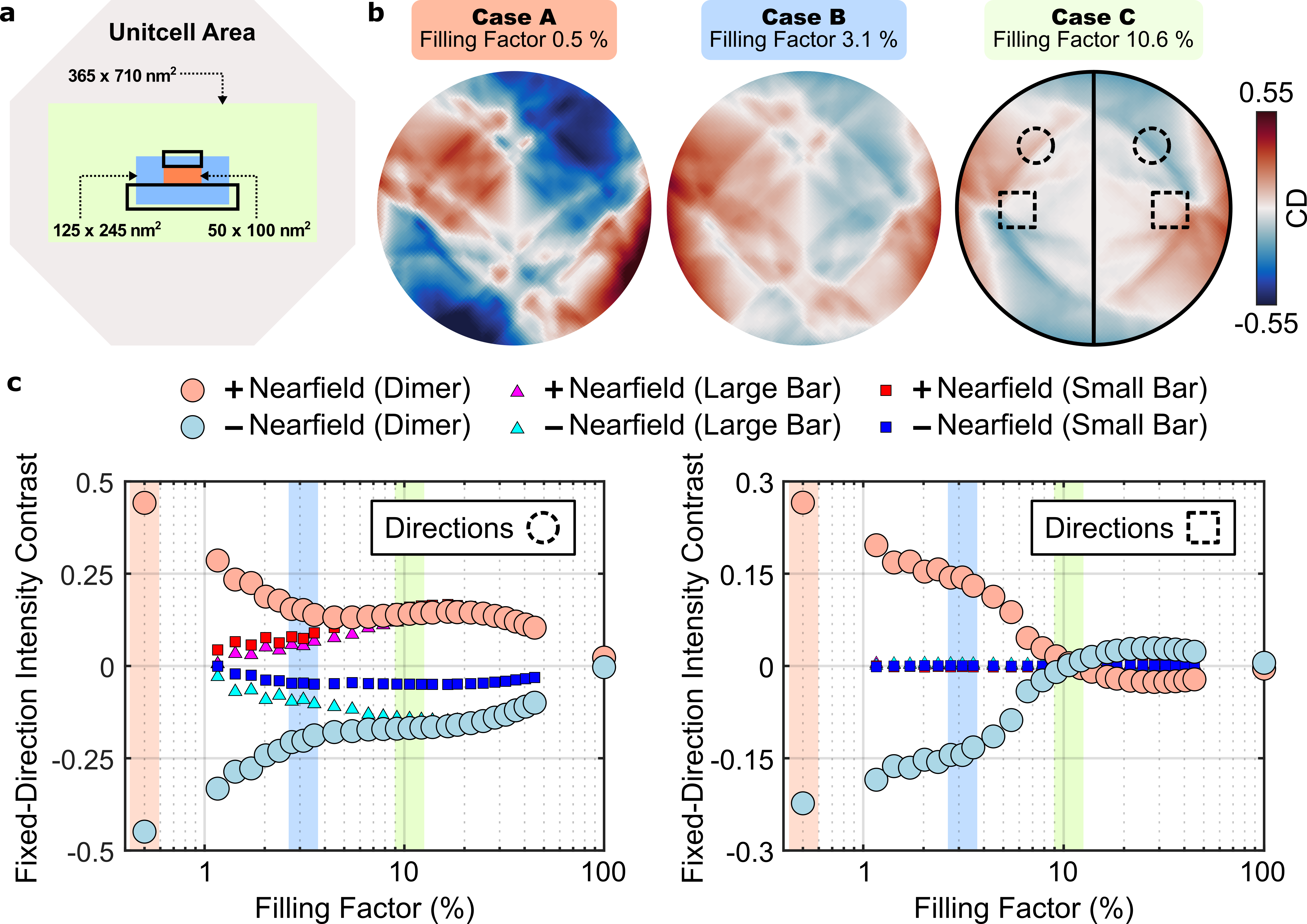}
    {\phantomsubcaption\label{subfig:numerics-nearfieldarea_a}}
    {\phantomsubcaption\label{subfig:numerics-nearfieldarea_b}}
    {\phantomsubcaption\label{subfig:numerics-nearfieldarea_c}}
	\caption{\textbf{Effect of the emitter distribution.} (\subref{subfig:numerics-nearfieldarea_a}) Top-view sketch of differently sized monolayers (orange, blue, and green) covering the unitcell area (grey) of the nanoantenna array. (\subref{subfig:numerics-nearfieldarea_b}) Numerically calculated angular circular dichroism of the nanoantenna array for different monolayer areas. Here, the spatial directions align with the nanoantenna sketch shown in (\subref{subfig:numerics-nearfieldarea_a}) and the angular range coincides with the experimental NA of 0.9. (\subref{subfig:numerics-nearfieldarea_c}) Retrieved directional intensity contrast as a function of the monolayer filling factor obtained for the directions indicated by the dashed circles (left) and dashed squares (right). Valley-selective results are obtained respectively from $\sigma^+$ and $\sigma^-$ polarized nearfield components for the nanoantenna array (red and blue circles), as well as for arrays of the individual large nanobar (magenta and cyan triangles) and the small nanobar (red and blue squares).}
	\label{fig:numerics-nearfieldarea}
\end{figure}

We recall that both the experimentally and numerically observed angular CD remain systematically below values predicted for valley-emitters placed in the vicinity of a single nanoantenna.~\cite{Chen2018routing} This deviation likely stems from contributions of emitters located farther away from the nanoantenna, experiencing weaker nearfield interactions. A natural way of testing this hypothesis is to adjust the monolayer area in the numerical simulations, effectively reducing the contributions of these remote emitters. Here, we numerically explore the potential of varying the monolayer extent within the unitcell, as a free design parameter, leaving other sample parameters intact.\\
Exemplary, we discuss several cases of differently sized monolayers placed on top of the nanoantenna within each unitcell. \autoref{subfig:numerics-nearfieldarea_a} and~\autoref{subfig:numerics-nearfieldarea_b} show the monolayer areas and, respectively, the calculated angular CD distributions for three different cases (A-C).
For case A, i.e. the monolayer filling the gap region between the nanobars, the angular CD takes magnitudes up to $53\%$. As the monolayer area is further increased (cases B and C), the range of the angular CD drops significantly due to the emission of mostly uncoupled emitters, exhibiting no valley-selective emission patterns. Additionally, this results in a qualitative change of the angular CD pattern, as clearly visible when comparing cases A and C. The pattern in A mostly reflects the nanoantenna's antisymmetry with respect to the $k_y$ axis (see also~\autoref{subfig:cryoPL-BFP_b}), due to the efficient nearfield interaction of emitters located within the nanoantenna's gap region. In C, most of the emitters are located further apart from the nanoantenna, and we find a two-fold antisymmetric distribution that originates from the square array arrangement.\\
Based on these considerations, we propose the following metric to quantitatively characterize the strength of the valley-routing effect. We consider two farfield detectors measuring the intensities, $I_\mathrm{left}$ and $I_\mathrm{right}$, emitted into two fixed farfield directions within the left and right halfspaces, respectively. Initially, we chose the two farfield directions indicated by the dashed circles in~\autoref{subfig:numerics-nearfieldarea_b}. We then define a measurable farfield quantity as the intensity contrast $(I_\mathrm{left}-I_\mathrm{right})/(I_\mathrm{left}+I_\mathrm{right})$ between both directions. In the left panel in~\autoref{subfig:numerics-nearfieldarea_c}, we analyze this directional intensity contrast obtained for different filling factors of the monolayer flake covering the nanoantenna array's unitcell (red and blue circles). For a fixed valley state or nearfield polarization, this contrast yields a finite quantity, with larger contrast values being obtained for smaller filling factors. Importantly, the sign of the contrast follows the valley-state, allowing the efficient readout of the valley-information from the farfield even in the presence of a resonant nanoscatterer, often obscuring this link due to its scattering response.~\cite{Bucher2024influence}\\
For comparison, we also calculated the intensity contrast obtained for arrays of individual nanobars, where we have adopted the same geometric parameters of the small nanobar (red and blue squares) and the large nanobar (magenta and cyan triangles). For small filling factors, we observe a clear difference in their directional scattering behavior compared to the nanobar dimer antenna. Hence, we conclude that the large directional contrast values for filling factors below $\sim 3\%$ result from the resonant interaction between both nanobars. For larger filling factors, however, the difference between dimer and monomer arrays becomes negligible, as the directional scattering is mostly influenced by diffractive modes.\\
Alternatively, in the right panel in~\autoref{subfig:numerics-nearfieldarea_c} we calculated the directional contrast for two different directions as indicated by the dashed squares in~\autoref{subfig:numerics-nearfieldarea_b}. Different to the previous case, we now find a clear difference in the directional contrast exhibited by the nanobar dimer array compared to the monomer arrays for all filling factors. Thus, we conclude that the directional contrast for this pair of farfield directions purely results from the valley-dependent directional scattering of the nanoantenna, with no additional contributions from diffractive modes.\\
It is important to note that the two previous cases reveal two different symmetries within the system, namely that of the lattice and that of the dimer nanoantenna. Importantly, only for the dimer nanoantenna this directional intensity contrast between left and right halfspaces is a symmetry-protected property, owing to the extrinsically induced chirality.~\cite{Hentschel.2012l1,Plum.2016,Caloz.2020,Sperrhake.2021} For a detailed symmetry analysis, see Sec.~S.7 of the Supporting Information. Consequently, when integrating over all farfield directions within each of the halfspaces separately, the directional intensity contrast mediated by the nanoantenna dimer is expected to be robust, while that for the arrays with single nanobars is expected to vanish. We have numerically verified this behavior as discussed in Sec.~S.8 of the Supporting Information. This symmetry-protected valley-dependent behavior therefore provides a robust platform for valleytronic processing owing to certain tolerances with respect to deviations in farfield directions and emitter distribution.
\section*{Conclusion}
We have experimentally investigated valley-momentum coupled emission from hybrid nanophotonic structures consisting of CVD-grown \ML{} placed on top of circular-polarization selective directional nanoantennas. Our hybrid nanostructures exhibit an antisymmetric angular CD with a magnitude of $6\%$. Crucially, no polarization analyzer is employed, clearly indicating the pivotal role of angle-selective coupling of the nanoantennas to valley-polarized exciton populations. Further, we have discussed the importance of experimental conditions able to distinguish valley-momentum locking from farfield-polarization sensitive directionality in the context of harnessing nanophotonics for valleytronics. These findings are supported by a novel numerical approach, allowing to model incoherent valley-selective emitters within periodic nanostructures. Our numerical results agree well with the experimental observation that the nanoantenna array exhibits a systematically lower angular CD than predicted for a single nanoantenna. By varying the coverage of the unitcell area by the monolayer, we are able to show that this discrepancy results from a reduced coupling of emitters distant to the nanoantenna. We demonstrate numerically that angular CD values above $50\%$ can be retained when limiting the monolayer area to the nanoantenna's gap region. Importantly, such capabilities may be realized in experiment using already existing techniques for nanopatterning of monolayer TMDs.~\cite{Mupparapu2020facile,Loechner2019controlling} Ultimately, we show that this directional intensity contrast can be harnessed robustly, owing to the extrinsically chiral properties of the dimer nanoantenna, providing a novel platform for valleytronic processing based on resonant nanoantennas.\\
Generally, the performance of nanoscopic valleytronic devices based on the non-vanishing DOVP will be challenged by two aspects: (1) The efficient optical addressing and reading out the valley degree of freedom in the presence of resonant nanostructures due to complex nearfield effects both at the excitation and emission level. Importantly, our scheme for valley-routing does not depend on the polarization state of emission in the farfield but relies on the valley-selective nearfield excitation alone. This provides a practical solution to overcome the limitation of an otherwise partly or fully obscured valley-information as measured in the farfield by encoding the valley-information in the PL emission direction. (2) Operating those devices at room temperature due to ultrafast spin-relaxation processes. Promisingly, several works demonstrate a non-zero DOVP at room temperature involving charge capturing~\cite{Hanbicki2016high}, alloying~\cite{Liu2020room-temperature}, multilayer~\cite{Gong2018nanowire} and heterostructure~\cite{Paradisanos2020prominent} materials paving the way for room temperature plasmonic and dielectric nanoantenna based valleytronic devices.
\section*{Methods}
\subsection*{Fabrication and nearfield characterization of gold nanoantennas}
Gold nanoantennas were fabricated on a thermally oxidized silicon wafer (\SI{300}{\nano\metre} silicon dioxide) using electron-beam lithography followed by a lift-off process. Initially, the electron-beam resist (ZEP520A) was spin-coated on the pre-cleaned silicon wafer at 4000\,rpm for 40\,s and baked at 180\,\textcelsius\, for 120\,s. Next, the nanoantenna arrays were defined by electron-beam lithography using a base dose of \qty{100}{\upmu\coulomb\per\centi\metre\squared} and dose factor variations between 1.2\textendash 1.6 and 1.4\textendash 1.8 for the large and small nanobars, respectively. After electron beam exposure, the resist was developed in N-amyl acetate for 90\,s and rinsed by isopropanol to remove the exposed areas of the resist. Subsequently, \qty{40}{\nano\metre} of gold were deposited using electron-beam evaporation (>99.5\,\% target purity, \qty{9e-6}{}\,Torr chamber pressure). For lift-off, the coated substrate was left for several hours in a N-N dimethyl acetamide bath followed by 10\,s ultrasonication and rinsing with acetone and isopropanol.\\ \\
\noindent Cathodoluminescence imaging was performed using a FEI Verios scanning electron microscope equipped with a Gatan MonoCL4 Elite detection system.
\subsection*{Optical experiments}
Angle-resolved white light spectroscopy was performed using a stabilized tungsten-halogen light source. The white light was prepared under circular polarization using a linear polarizer and a super-achromatic quarter wave plate and focused onto the sample by a 100x/0.88NA objective resulting in an illumination spot diameter of about $\SI{5.5}{\micro\metre}$. In reflection geometry, the back-scattered light was collected by the same objective, passed through a 30(R):70(T) plate beam splitter, and analyzed in wavelength-momentum space by imaging the back-focal plane of the objective onto the entrance slit of an imaging spectrometer (Andor Shamrock 750). This resulted in a spectral and momentum resolution of $\Delta\lambda=\SI{0.23}{\nano\metre}$ and $\Delta k_x/k_0 = 0.016$. The momentum-axis was defined by orienting the sample with respect to the orientation of the spectrometer slit. A sketch of the optical setup and a detailed discussion of the polarization control are provided in Sec.~S.1 of the Supporting Information.\\ \\
\noindent Room temperature photoluminescence measurements were conducted using a commercial confocal fluorescence lifetime imaging microscope (MicroTime 200, PicoQuant). A pulsed laser source ($\SI{100}{\pico\second}$ pulse length, $\SI{40}{\mega\hertz}$ repetition rate, $\SI{30}{\micro\watt}$ average power) at $\SI{532}{\nano\metre}$ wavelength was used for excitation and focused on the sample by a 100x/0.95NA objective resulting in an estimated spot diameter of $2r = 2\lambda/(\mathrm{NA}\cdot\pi)\approx\SI{0.36}{\micro\metre}$. The signal was collected by the same objective in reflection geometry. The reflected laser light was blocked using a \SI{550}{\nano\metre} longpass filter.\\ \\
\noindent Polarization-resolved cryogenic photoluminescence measurements were conducted at a temperature of 3.8\,K using a closed-loop helium cryostat (s50, Montana Instruments). A 633\,nm wavelength continuous-wave helium-neon laser ($\SI{100}{\micro\watt}$ average power) was used for excitation. The laser light was prepared under circular polarization using a linear polarizer and a quarter wave plate and focused on the sample by a 100x/0.88NA objective resulting in an estimated spot diameter of $2r = 2\lambda/(\mathrm{NA}\cdot\pi)\approx\SI{0.46}{\micro\metre}$. The signal was collected in reflection geometry by the same objective, passed through a dichroic beam splitter and analyzed in a helical polarization basis by a super-achromatic quarter wave plate and a linear polarizer. For angle-resolved PL imaging, the back-focal plane of the objective was directly imaged onto an electron-multiplying charge-coupled device (EMCCD, iXon897 Ultra, Andor) or onto the entrance slit of an imaging spectrometer as described above. On the EMCCD, this resulted in a momentum resolution of $\Delta k_x/k_0 = 0.011$. A sketch of the optical setup and a detailed discussion of the influence of the optical components on the detection polarization are provided in Sec.~S.1 of the Supporting Information.
\subsection*{Numerical simulations}
For modeling incoherent emission from periodic nanostructures, we employ an \textit{Averaged Reciprocal Modal Analysis} (ARMA), exploiting Lorentz reciprocity. This allows us to infer the emission pattern of a two-dimensional semiconductor from its plane wave excitation response.~\cite{Vaskin2019lightemitting,Pal2025dark}\\
We model this by scanning a plane wave over a regular grid of incident wavevectors $\mathbf{k}=(k_x,k_y)$, bound by the numerical aperture $|k|\leq NA = 0.9$, and computing the respective nearfield intensity distribution of the nanoantenna array at the emission wavelength and within the monolayer plane. By reciprocity, we approximate the polarization-dependent angular PL intensities measured in the experiment by the obtained spatially-averaged nearfield intensity.\\
Applying this concept to valley-selective emission, we additionally identify the circularly polarized components of the nearfield to emission contributions obtained from respective valley-excitons. Ultimately, we obtain a polarization-dependent spatially-averaged nearfield intensity in $k$-space, allowing qualitative analysis of emission directionality features connected to polarization and nearfield asymmetries.
\subsubsection*{Fourier modal method}
For the full-wave calculations, we employed a custom three-dimensional Fourier modal method (also known as rigorous coupled‐wave analysis) with two-dimensional in-plane periodicity.~\cite{Noponen.1994, Li.1996, Li.1997} The simulation domain is a $1\upmu\mathrm{m}\times 1\upmu\mathrm{m}$ unitcell containing the nanobar dimer and the layered dielectric stack (compare~\autoref{fig:sketch} and~\autoref{fig:sample}). The angular space of the incident plane waves is defined in $k$-space via
\begin{align}
	\begin{matrix}
		k_x &=& \sin(\theta)\cos(\phi),\\
		k_y &=& \sin(\theta)\sin(\phi)
	\end{matrix}
	\quad
	\Rightarrow
	\quad
	\begin{matrix}
		\theta &=& \arcsin(\sqrt{k_x^2 + k_y^2}),\\
		\phi &=& \arctan2(k_y,k_x),
	\end{matrix}
\end{align} 
where $\sqrt{k_x^2 + k_y^2}\le 0.9$. The numerical solutions obtained by the Fourier modal approach yield the scattering matrix of the layered system, whose entries are the complex amplitudes of all reflected and transmitted diffraction orders at the monolayer plane. The evanescent mode coefficients then comprise the nearfield components originating from the nanoantenna that interact with the adjacent layers.~\cite{Menzel.2010, Sperrhake.2021}\\
The electric field components in a linear polarization basis then follow from the Rayleigh-expansion,~\cite{Noponen.1994} as
\begin{equation}
	E_n(x,y,z) = \sum_{l,m=-N}^{N} \mathcal{E}_{n,lm} \, \exp\left(i \left[\gamma_{lm} z + k_{x,l} x + k_{y,m} y\right]\right), \label{eq:rayleighseries}
\end{equation}
where \( \mathcal{E}_{n,lm} \) are the Rayleigh coefficients of the $n-$th field component, with $n\in \{x,y,z\}$, obtained from the scattering matrix coefficients $(l,m)$. The propagation constant in $z$-direction is given by $\gamma_{lm} = \sqrt{\left(\tilde{n}k_0\right)^2 - k_{x,l}^2 - k_{y,m}^2}$, with the vacuum wavenumber $k_0$ and the refractive index of the surrounding medium $\tilde{n}$.
\subsubsection*{Helical basis projection}
To impose circular polarization, we first project the incident field from its $s$-$p$ basis~\cite{Noponen.1994} in spherical coordinates onto the in-plane Cartesian coordinate system of the monolayer. For a given emission direction $\mathbf{k} = (\sin\theta\cos\phi,\, \sin\theta\sin\phi,\, \cos\theta)$, we find:
\begin{align}
	\mathbf{p} &= (\cos\theta\cos\phi,\; \cos\theta\sin\phi,\; -\sin\theta), \\
	 \mathbf{s} &= (-\sin\phi,\; \cos\phi,\; 0),
\end{align}
with the respective transformation matrix:
\begin{equation}
	U = 
	\begin{bmatrix}
		\mathbf{s}\\
		\mathbf{p}
	\end{bmatrix}\cdot
	[\mathbf{x} \, \mathbf{y}] = 
	\begin{pmatrix}
		\cos\theta\cos\phi & \cos\theta\sin\phi \\
		-\sin\phi & \cos\phi
	\end{pmatrix}.
\end{equation}
For oblique emission angles, any circularly polarized farfield component yields an elliptical projection onto the monolayer plane, with the ellipticity and orientation of the polarization ellipse being determined by the emission angle. Hence, we find the in-plane nearfield distribution $(E_{\pm,x}^\prime, E_{\pm,y}^\prime)^T$ upon a $\sigma^\pm$ polarized plane wave, as:
\begin{equation}
    \begin{pmatrix}
    E_{\pm,x}^\prime\\
    E_{\pm,y}^\prime
    \end{pmatrix}
    =
    \frac{1}{\sqrt{2}}\, U 
    \begin{pmatrix} 
    E_x \\ 
    \pm i\cdot E_y
    \end{pmatrix} 
	= \frac{1}{\sqrt{2}} \begin{pmatrix}
		\cos\theta\, e^{i\phi}\cdot E_x \\
		(\mp\sin\phi \pm i\,\cos\phi)\cdot E_y
	\end{pmatrix}.
\end{equation}
Finally, we transform the obtained in-plane nearfield distribution from a linear polarization basis to a helical basis,~\cite{Menzel.2010} as:
\begin{equation}
	\begin{pmatrix}
		E_{+,\pm} \\
		E_{-,\pm}
	\end{pmatrix}
	= \frac{1}{\sqrt{2}}
	\begin{pmatrix}
		1 & 1 \\
		i & -1
	\end{pmatrix}
	\begin{pmatrix}
		E_{\pm,x}^\prime \\
		E_{\pm,y}^\prime
	\end{pmatrix},
\end{equation}
with the Jones notation $\sigma^+_{\mathrm{farfield}}|\sigma^-_{\mathrm{nearfield}}\equiv+-$. In this helical bases, we are able to compare the emerging patterns in $k$-space from the ARMA approach to the measured valley-selective angular PL intensities.
\subsubsection*{Field averaging and box-size dependence}
In order to obtain the contribution of the nearfield intensities, reciprocally yielding emission contributions in different angular directions, we calculate the averaged nearfield intensity related to that plane wave direction
\begin{equation}
	M^{\pm}(k_x,k_y;\{x,y \in \text{box}\}) = \frac{1}{N_{\text{box}}}\sum_{\{x,y \in \text{box}\}} |E_{\sigma^{\pm}}(k_x,k_y;x,y)|^2.
\end{equation}
We mimic the experimental conditions by choosing an integration box covering the entire unitcell area.\\
Additionally, by varying the averaging box size, we are able to investigate both spatially- and polarization-dependent directional coupling effects between valley-selective emitters and the nanobar dimer antennas.
%
%

\begin{acknowledgement}

We thank A. Lamprianidis for insightful discussions on the helical symmetry properties of the studied system. This work was funded by the Deutsche Forschungsgemeinschaft (DFG, German Research Foundation), Project-IDs: 398816777 -- SFB 1375 (NOA), 437527638 -- IRTG 2675 (Meta-Active) and 448835038. Further DFG funding was received under Germany's Excellence Strategy – EXC 2051 – Project-ID 390713860 and Affiliation: Cluster of Excellence Balance of the Microverse, Friedrich Schiller University Jena, Jena, Germany. We also acknowledge financial support by the State of Thuringia (Quantum Hub Thüringen, 2101 GFI 0043, Qi2.7) and by the Australian Research Council Centres of Excellence Program (CE200100010). This work was partly performed at the ACT node of the Australian National Fabrication Facility (ANFF-ACT). 

\end{acknowledgement}

\begin{suppinfo}

Supporting Information Available: (Sec.~S.1) Sketch of the optical setup and characterization of the polarizing properties of the optical components. (Sec.~S.2) Diffractive grating orders and grating order dispersion in momentum-space. (Sec.~S.3) Room temperature photoluminescence characterization of \ML{} placed on top of a fabricated gold nanoantenna array. (Sec.~S.4) Spectral analysis of the excitonic PL from \ML{}. (Sec.~S.5) Comparison of numerical and experimental CD values. (Sec.~S.6) Numerically calculated angular emission patterns. (Sec.~S.7) Symmetry properties and extrinsic chirality of the nanoantenna array. (Sec.~S.8) Angle-integrated farfield intensity contrast. (Sec.~S.9) Farfield circular polarization contrast. (PDF)

\end{suppinfo}


\bibliography{references}

\end{document}


\setcounter{page}{1}
%
%
\section{Influence of the optical components on the\\ polarization state of light}
\begin{figure}[h]
	\centering
	\includegraphics[width=15cm]{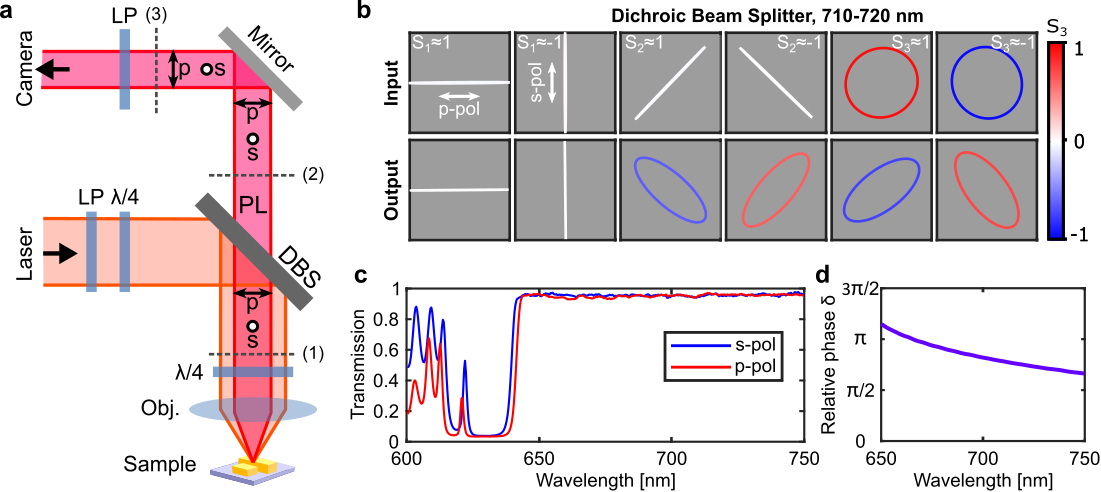}
    {\phantomsubcaption\label{subfig:DBS_a}}
    {\phantomsubcaption\label{subfig:DBS_b}}
    {\phantomsubcaption\label{subfig:DBS_c}}
    {\phantomsubcaption\label{subfig:DBS_d}}
	\caption{\textbf{Optical setup and polarization control.} (a) Sketch of the experimental setup. The numbers show different positions of the setup where the polarization state can be examined. (b) The effect of the dichroic beam splitter on the polarization state of light in transmission. The upper row of the table shows the polarization ellipses of the incoming light. The bottom row are the corresponding polarization ellipses being modified by the DBS normalized to the incoming intensity. The measurements are averaged over the spectral range from 710 nm to 720 nm, the colorcode of the ellipses represents their $S_3$ value. (c) Wavelength-dependent transmission of the DBS for s- and p- polarized light. (d) The relative phase shift between s- and p- components induced by the DBS.}
	\label{fig:SI_DBS}
\end{figure}

We performed optical experiments in reflection mode as illustrated in~\autoref{subfig:DBS_a} and~\autoref{subfig:SI-setup-kspace_a}. In these configurations, the incoming light is initially reflected by a beam splitter (see position (1)) and focused onto the sample inside the cryostat via an objective lens. The circular polarization state of the excitation/illumination is routinely checked before entering the objective. The light emitted from the sample is collected by the same objective, transmitted through the beam splitter (see position (2)), and redirected by a mirror (see position (3)). In the following, we examine the influence of the beam splitter and the mirror on the measured polarization state of the light collected in this configuration. For this, we use the Stokes-formalism of describing (partially) polarized light. The normalized Stokes vector ${\textbf{S}} = \left(1, S_1, S_2, S_3\right)$ is then defined as follows:
    \begin{align}~\label{eq:Stokes_Parameters}
        S_1 = \frac{I_\mathrm{s} - I_\mathrm{p}}{I_\mathrm{s} + I_\mathrm{p}},\hspace{0.4cm}S_2 = \frac{I_\mathrm{d} - I_\mathrm{a}}{I_\mathrm{d} + I_\mathrm{a}},\hspace{0.4cm}S_3 = \frac{I_{\sigma^+} - I_{\sigma^-}}{I_{\sigma^+} + I_{\sigma^-}},
    \end{align} 
where $I$ denotes the light intensities measured for different polarization configurations in detection, as indicated by their subscripts:  linear vertical (s) and horizontal (p), linear diagonal (d) and antidiagonal (a), as well as left-handed ($\sigma^{+}$) and right-handed ($\sigma^{-}$) circular polarization.
\subsection{Circular polarization resolved measurements}
\subsubsection{Scheme for polarization control}
For circular-polarization resolved measurements (see Fig. 4(b,c) and Fig. 5 in the main text) we have used a longpass dichroic beam splitter (DBS) with a cut-on wavelength of \SI{650}{\nano\metre} as shown in~\autoref{subfig:DBS_a}. As DBS are known to change the polarization state of elliptically polarized light, we have employed the following scheme: We used a combination of a linear polarizer (LP) and a quarter-wave plate (QWP) as compensating elements to prepare s-polarized light after the reflection from the DBS (position (1)). The s-polarized light was then sent through a super-achromatic QWP in order to prepare $\sigma^\pm$ polarized light before entering the objective. In detection, the same super-achromatic QWP is used to convert the circular polarized components of the collected light into a linear basis and the s- and p-polarized light is transmitted through the DBS (position (2)), reflected by a mirror (position (3)) and analyzed by a LP.

\subsubsection{Dichroic beam splitter}
We examined the polarizing effect of the DBS on the transmitted light in a custom-built white-light spectroscopy setup designed for near-zeroth order transmittance (NA$\approx 0.044$) at a 45$^\circ$ incidence angle. In this configuration, incoming white light, prepared in an arbitrary polarization state, passes through the DBS and is fiber-coupled to a spectrometer. A QWP in a motorized rotation mount and a fixed LP are positioned before the spectrometer fiber to measure the Stokes vector components of the transmitted light using the Fourier method. We characterized the transmitted light across six degenerate input polarization states as defined for the Stokes parameters above. \autoref{subfig:DBS_b} shows the polarization ellipses of the incident (upper row) and transmitted (lower row) light averaged over a narrow spectral range of \SIrange{710}{720}{\nano\metre}. We observe that: (1) s- and p-polarized states remain almost unchanged, (2) diagonal linear polarization turns into elliptical, while the circular states become elliptical as well and their $S_3$ parameter changes sign. From these observations we can conclude that the DBS induces a phase shift $\delta$ between s- and p-polarized field components.\\
\autoref{subfig:DBS_c} and~\autoref{subfig:DBS_d} show the measured transmittance and relative phase shift spectra of the DBS, respectively, where $T_{s/p}=I_{\mathrm{s/p}}^\mathrm{out}/I_{\mathrm{s/p}}^\mathrm{in}$ and the relative phase shift $\delta$ between s- and p-polarized components was extracted from analyzing how the diagonal and circular-polarized components were modified by the DBS, namely, $\delta=\phi^\mathrm{in}-\phi^\mathrm{out}$, where $\phi^\mathrm{in/out}=\mathrm{tan}^{-1}\left(\frac{S^\mathrm{in/out}_3}{S^\mathrm{in/out}_2}\right)$. The phase shift was extracted as an average over four measurement sets (two diagonal and two circular states were analysed) with a standard deviation of $<0.012$ rad. For the wavelength range from \SIrange{690}{750}{\nano\metre} we find that indeed the amplitude modulation of the DBS is negligible ($T_s\approx T_p\approx 0.96=T$) but a significant phase shift is introduced.\\  
Next, to characterize the polarization behavior of the DBS, we utilize the Müller matrix formalism. This approach relates the output Stokes vector $\textbf{S}^{\mathrm{out}}$ to the input Stokes vector $\textbf{S}^{\mathrm{in}}$ by means of a 4x4 matrix $\hat{\mathrm{M}}_{\mathrm{DBS}}$ such that $\textbf{S}^{\mathrm{out}} = \hat{\mathrm{M}}_{\mathrm{DBS}}\textbf{S}^{in}$. Given our observations, the DBS acts as a waveplate and consequently, its Müller matrix takes the following form:
\begin{equation}
        \hat{\mathrm{M}}_{\mathrm{DBS}}(\lambda)  =T\begin{bmatrix}
            1 & 0 & 0 & 0\\
            0 & 1 & 0 & 0\\
            0 & 0 & \cos{\delta(\lambda)} & -\sin{\delta(\lambda)}\\
            0 & 0 & \sin{\delta(\lambda)} & \cos{\delta(\lambda)}
        \end{bmatrix}.
    \end{equation}
\subsubsection{Mirror}
    After the DBS, the light is reflected from a protected silver mirror under 45$^{\circ}$. The Müller matrix for reflection from metals can be given in terms of the complex reflection coefficients $r_s$ and $r_p$
    \begin{equation}
        M_{\mathrm{Mirror}}  =\frac{1}{2}\begin{bmatrix}
            r_s^2+r_p^2 & r_s^2-r_p^2 & 0 & 0\\
            r_s^2 -r_p^2 & r_s^2+r_p^2 & 0 & 0\\
            0 & 0 & 2r_s r_p\cos{\gamma} & -2r_s r_p\sin{\gamma}\\
            0 & 0 & 2r_s r_p\sin{\gamma} & 2r_s r_p\cos{\gamma}
        \end{bmatrix},
    \end{equation}
    where $\gamma=\phi_s-\phi_p$ is the phase offset between s- and p- polarized field components. We assume that this phase shift is approximately constant through the examined wavelength range. According to characterization by the manufacturer (Thorlabs), the reflectance of the protected silver mirror at \SI{710}{\nano\metre} wavelength and for 45$^{\circ}$ incidence angle are $r_s^2=R_s\approx0.96$ and $r_p^2=R_p\approx0.95$. Hence, we approximate that $R_s\approx R_p\equiv R$ with reasonable accuracy.

%
%
\subsection{Angle-resolved spectroscopy}
\subsubsection{Scheme for polarization control}
\begin{figure}[h]
	\centering
	\includegraphics[width=15.5cm]{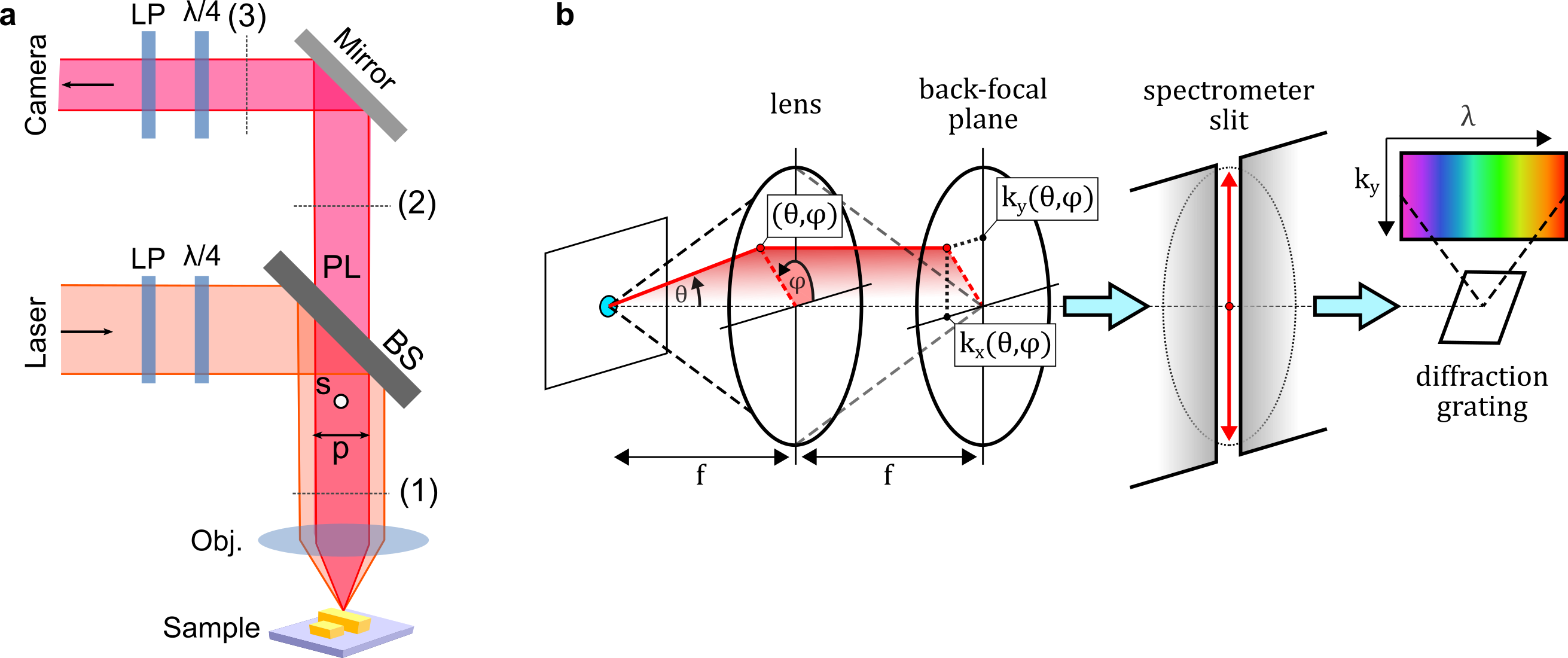}
    {\phantomsubcaption\label{subfig:SI-setup-kspace_a}}
    {\phantomsubcaption\label{subfig:SI-setup-kspace_b}}
	\caption{\textbf{Angle-resolved spectroscopy setup.} (\subref{subfig:SI-setup-kspace_a}) Sketch of the experimental setup configuration used for angle-resolved spectroscopy. (\subref{subfig:SI-setup-kspace_b}) Sketch of the imaging setup used for angle-resolved spectroscopy.}
	\label{fig:SI-setup-kspace}
\end{figure}

For angle-resolved spectroscopy measurements (see Fig. 2(d) and Fig. 4(d-g) in the main text) we have used the setup configuration as shown in~\autoref{subfig:SI-setup-kspace_a} with a 30(R):70(T) plate beam splitter (BS). As plate BS are known to change the polarization state of elliptically polarized light, we have employed the following scheme: We used a combination of a LP and a QWP as compensating elements to prepare circular polarized light after the reflection from the BS (position (1))and before entering the objective. In detection, the circular polarized components of the collected light are transmitted through the BS (position (2)), reflected by a mirror (position (3)) and analyzed by a super-achromatic QWP and a LP.
\subsubsection{30(T):70(R) plate beam splitter}
We have provided a detailed characterization of the polarizing properties of the BS in Sec.~S.1 of the Supporting Information of a previous work.~\cite{Bucher2024influence} In short, the 30(T):70(R) beam splitter acts as a linear polarizing element with anisotropic amplitude attenuation coefficients $p_s$ and $p_p$, where $T_s = p^2_s$ and $T_p = p^2_p$ are the transmittances for $s$- and $p$-polarization, respectively. The measured transmittance spectra of the beam splitter showed a minute wavelength dependence and an average transmittance of $T_s=0.86$ and $T_p=0.61$ in the wavelength range of \SIrange{650}{750}{\nano\metre}. If we consider a partially circular-polarized input state, represented by its Stokes vector, $\textbf{S}^{\mathrm{in}}=\left[1,0,0,\alpha\right]^\mathrm{T}$ where $\alpha\in(-1,1)$, the light transmitted through the beam splitter would be described by $\textbf{S}^{\mathrm{out}}=\hat{\mathrm{M}}_{\mathrm{BS}}\textbf{S}^{in}=0.735\cdot\left[1,0.170,0,0.985\cdot\alpha\right]^\mathrm{T}$ as previously shown. Therefore, the degree of circular polarization (DOCP) or $S_3$ of the light collected by the objective is slightly decreased by a factor of 0.985 after passing through the beam splitter which is comparable to the natural fluctuations occurring for measurements upon repetition or different sample positions.

\subsubsection{Back-focal plane imaging}
In order to resolve the collected light in angular-space, we have imaged the back-focal plane (BFP) of the objective onto the entrance slit of an imaging spectrometer (Shamrock 750, Andor) as sketched in~\autoref{subfig:SI-setup-kspace_b}. We have used two consecutive 4f imaging arms in order to image the BFP of the objective onto the spectrometer slit such that $k=0$ is falling onto the center of the slit. By closing the spectrometer slit ($\approx \SI{150}{\micro\metre}$), we isolate a narrow slice of the BFP (including $k=0$) whose orientation is defined by rotating the sample. The slice of the BFP which is transmitted through the slit is then collected by a parabolic silver mirror inside the spectrometer, sent onto a 1D diffraction grating (150~lmm$^-1$) and imaged onto a charge-coupled device (CCD) camera (iDus420, Andor). This resulted in a spectral and momentum resolution of $\Delta\lambda=\SI{1.4}{\nano\metre}$ and $\Delta k_x/k_0 = 0.016$.
\section{Diffractive grating orders in momentum space}
\begin{figure}[h!]
	\centering
	\includegraphics[width=10cm]{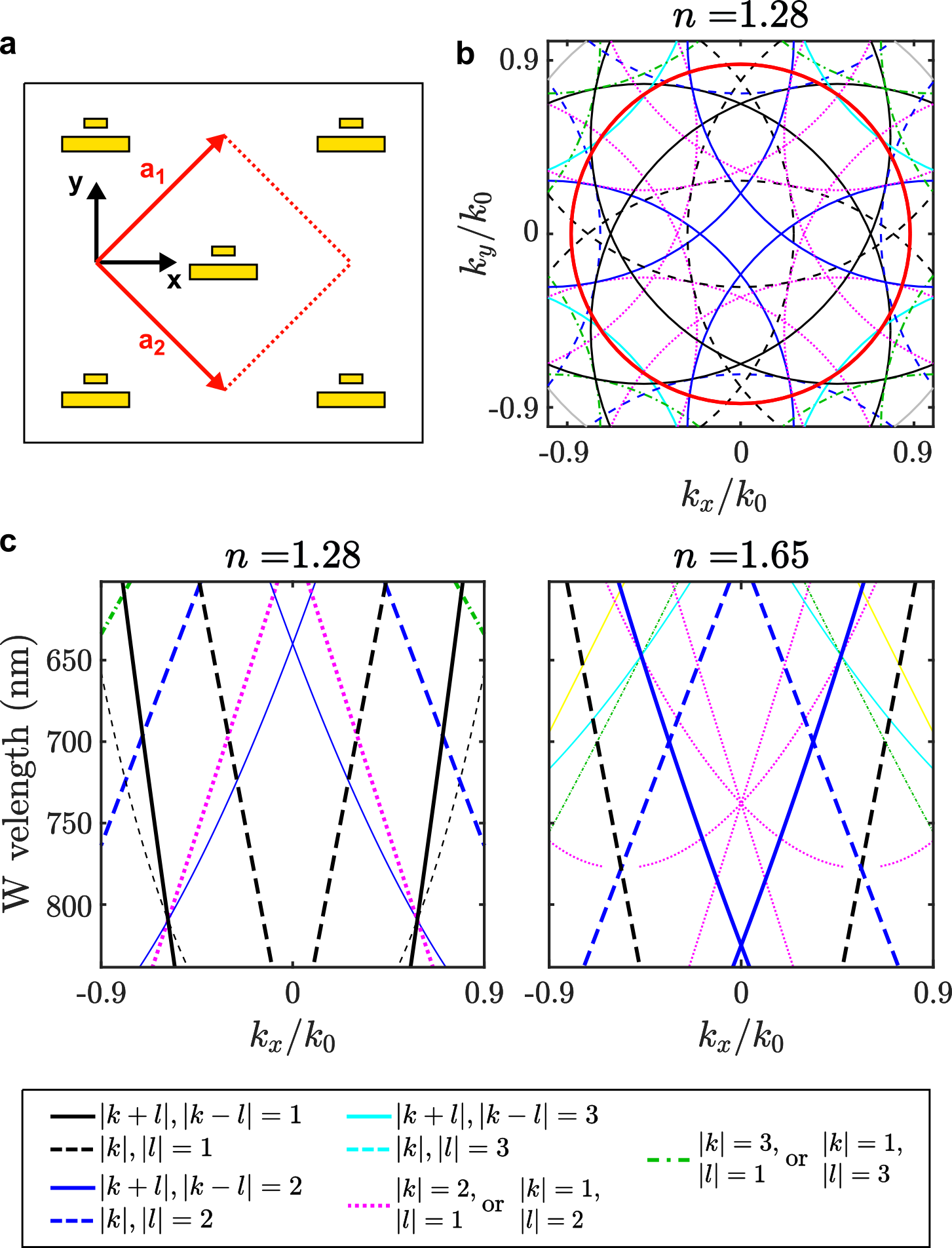}
    {\phantomsubcaption\label{subfig:SI-difforders_a}}
    {\phantomsubcaption\label{subfig:SI-difforders_b}}
    {\phantomsubcaption\label{subfig:SI-difforders_c}}
	\caption{\textbf{Diffractive modes of the nanoantenna array.} (\subref{subfig:SI-difforders_a}) Sketch of the nanoantenna array and the primitive unit cell of the square lattice (red lines). (\subref{subfig:SI-difforders_b}) Diffractive grating orders of the square array shown on the left with a lattice constant of $|\textbf{a}_1|=|\textbf{a}_2|=1\SI{1}{\micro\metre}$ and a refractive index of $n_1=1.28$.  The experimental numerical aperture is indicated by a red circle. (\subref{subfig:SI-difforders_c}) Grating order dispersion for the same grating as discussed above for $k_y=0$ considering refractive indices of $n_1=1.28$ (left) and $n_2=1.65$ (right). In all cases the lowest eight grating orders are highlighted by different colors according to the legend at the bottom.}
	\label{fig:SI-difforders}
\end{figure}

We have analyzed the influence of the diffractive grating orders exhibited by the periodic nanoantenna array (square lattice period $\Lambda=\SI{1}{\micro\metre}$) in momentum space. \autoref{subfig:SI-difforders_a} shows a sketch of the nanoantenna array and the primitive unit cell of the square lattice (red lines). We then find the respective (reciprocal) lattice vectors, expressed in the $xy$-coordinate system (black arrows), as
\begin{align}
    \text{real space:}&\hspace{0.3cm}\textbf{a}_1=\frac{\Lambda}{\sqrt{2}}\begin{pmatrix} 1\\ 1\end{pmatrix}, \hspace{0.3cm}\textbf{a}_2=\frac{\Lambda}{\sqrt{2}}\begin{pmatrix} 1\\ -1\end{pmatrix}\hspace{0.3cm}\text{and}\\
    \text{reciprocal space:}&\hspace{0.3cm}\textbf{b}_1=\frac{\sqrt{2}\pi}{\Lambda}\begin{pmatrix} 1\\ 1\end{pmatrix}, \hspace{0.3cm}\textbf{b}_2=\frac{\sqrt{2}\pi}{\Lambda}\begin{pmatrix} 1\\ -1\end{pmatrix}.
\end{align}
The linear scattering of a plane wave with incidence angles $(\theta,\phi)$, where $\theta\in[0,\pi]$ is the azimuthal angle measured from the direction of the substrate normal and $\phi\in[0,2\pi)$ is the polar angle, follows
\begin{align}
    \textbf{k}'_{||} = \textbf{k}_{||} + \textbf{G}_{kl}\hspace{0.15cm},\hspace{0.3cm}\text{with}\hspace{0.3cm}\textbf{G}_{kl}=k\textbf{b}_1 + l\textbf{b}_2\hspace{0.15cm},\hspace{0.3cm}k,l\in\mathbb{Z}
\end{align}
where $\textbf{k}_{||}=k_0(\cos{\varphi}\sin{\theta},\sin{\varphi}\sin{\theta})^{\text{T}}$ and $\textbf{k}'_{||}$ are the in-plane wavevectors of the incident and scattered wave, respectively, and $k_0 = 2\pi/\lambda$ is the wavenumber at wavelength $\lambda$. In the farfield, we only observe scattered waves if their out-of-plane component $k'_z$ is real or equivalently
\begin{align}
		\label{eq:difforders}
		n^2k_z^{'2} = n^2k_0^2 - \left(k_0\cos\varphi\sin\theta+(k+l)\frac{\sqrt{2}\pi}{\Lambda}\right)^2 - \left(k_0\sin\varphi\sin\theta+(k-l)\frac{\sqrt{2}\pi}{\Lambda}\right)^2\geq0,
\end{align}
where $n$ is the refractive index of the medium. For $k'_z=0$,~\autoref{eq:difforders} describes a set of circles in momentum space with radius $nk_0$ and center points $-\frac{\sqrt{2}\pi}{\Lambda}((k+l),(k-l))$. \autoref{subfig:SI-difforders_b} shows the respective solutions for $k'_z=0$ in momentum space for an refractive index of $n=1.28$.
We highlighted eight sets of grating orders which overlap with the k-space within the experimental numerical aperture (NA=0.88) by different color and line style. Here, orders with $|k+l|=|k-l|$ relate to modes propagating along the nearest neighbor direction (i.e. along the direction of the (inverse) lattice vectors) and orders with $|k|=|l|$ relate to modes propagating along the second-nearest neighbor direction (i.e. along the $x-$ and $y-$direction). Note that we have used the refractive index $n$ as a free parameter to fit the experimentally observed patterns in the back-focal plane images of PL from \ML{} on top of the nanoantenna array (see Fig. 5 in the main text). The nanoantennas are embedded in a 15\,nm thin layer of silicon dioxide. The low effective refractive index of $n_1=1.28$ might hint at a non-negligible porosity of the deposited thin film with pore sizes smaller than the optical wavelength.\\
By dividing~\autoref{eq:difforders} with $k_0^2$, we can rewrite the grating order equation at $k_x=0$ (i.e. $\varphi=0$) to obtain the implicit grating order dispersion relation
\begin{align}
		\label{eq:difforders-dispersion}
		n^2 - \left(\sin\theta+(k+l)\frac{\lambda}{\sqrt{2}\Lambda}\right)^2 - (k-l)^2\left(\frac{\lambda}{\sqrt{2}\Lambda}\right)^2\geq0,
\end{align}
where $\lambda$ is the vacuum wavelength of light. From~\autoref{eq:difforders-dispersion}, we find that the orders generally show a quadratic dispersion except for $k=l$ (i.e. the modes exhibited by a 1D grating along $x$-direction). \autoref{subfig:SI-difforders_c} shows the respective solutions for $k'_z=0$ in momentum space for refractive indices of $n_1=1.28$ (left side) and $n_2=1.65$ (right side). Again, we highlighted eight sets of grating orders which overlap with the k-space within the experimental numerical aperture (NA=0.88) by different color and line style. Note that in Fig. 2d of the main text only those orders are considered which show in the measured momentum-resolved spectra.
\section{Room temperature photoluminescence}
\begin{figure}[h!]
	\centering
	\includegraphics[width=15.5cm]{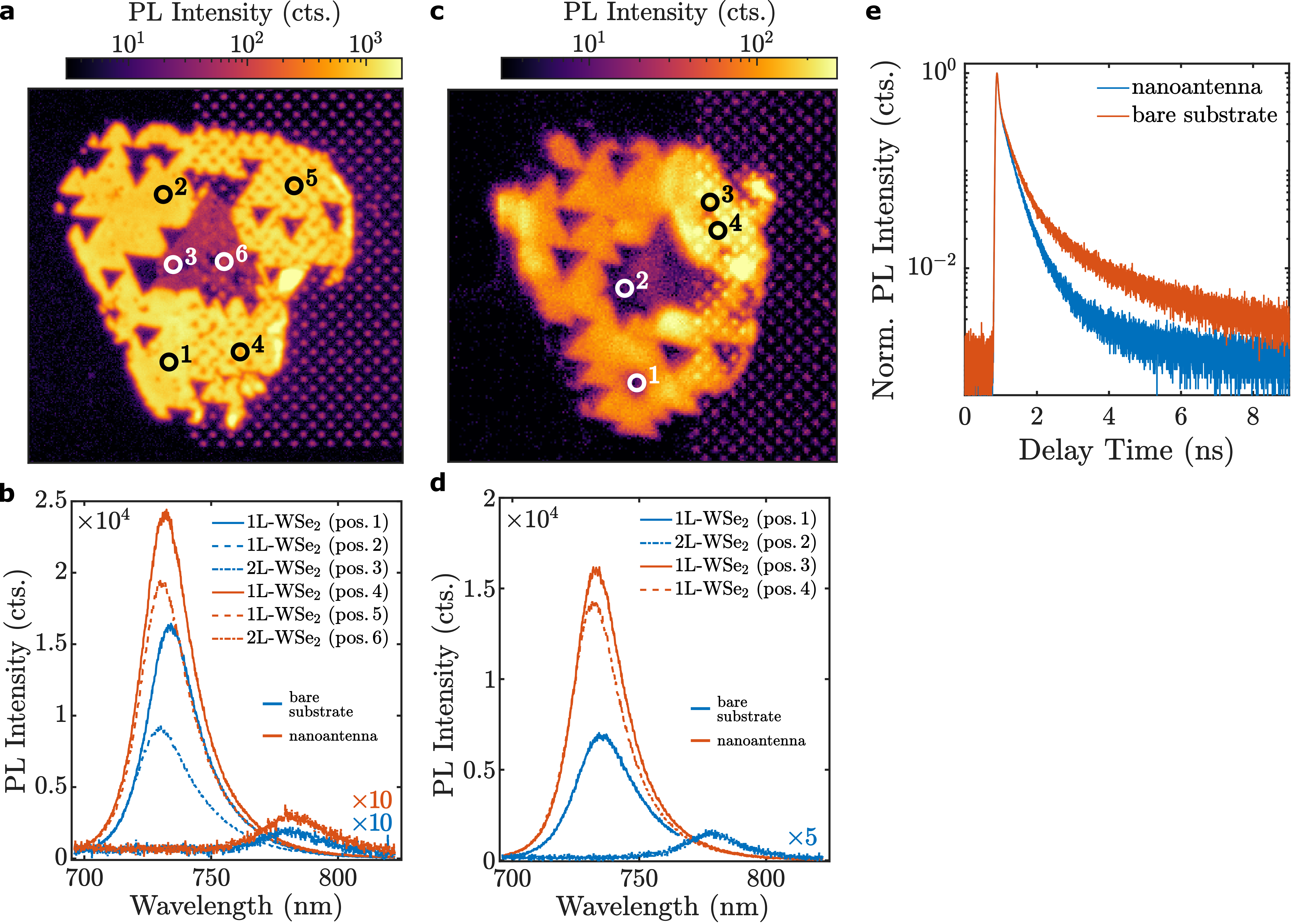}
    {\phantomsubcaption\label{subfig:rtpl_a}}
    {\phantomsubcaption\label{subfig:rtpl_b}}
    {\phantomsubcaption\label{subfig:rtpl_c}}
    {\phantomsubcaption\label{subfig:rtpl_d}}
    {\phantomsubcaption\label{subfig:rtpl_e}}
	\caption{\textbf{Additional room temperature PL measurements.} (\subref{subfig:rtpl_a},\subref{subfig:rtpl_c}) Measured confocal microscope photoluminescence images from \ML{} on top of nanoantenna arrays with different geometrical nanoantenna parameters. (\subref{subfig:rtpl_b},\subref{subfig:rtpl_d}) Respective photoluminescence intensity spectra measured in the positions indicated by circles in (\subref{subfig:rtpl_a},\subref{subfig:rtpl_c}) and labelled by numbers. (\subref{subfig:rtpl_e}) Measured photoluminescence decay curves of the same WSe\textsubscript{2} crystal as shown in (\subref{subfig:rtpl_a}) on top of the gold nanoantenna array (orange curve) and on bare substrate (blue curve).}
	\label{fig:SI-rt-pl}
\end{figure}

We characterized the hybrid system of monolayer WSe\textsubscript{2} on top of a gold nanoantenna array at room temperature by means of photoluminescence (PL) microscopy and spectroscopy. All measurements at room temperature were performed using a commercial fluorescence lifetime imaging setup (PicoQuant, MicroTime 200). A \SI{530}{\nano\metre} pulsed excitation laser with a \SI{40}{\mega\hertz} repetition rate, $\SI{100}{\pico\second}$ pulse duration, and $\approx \SI{30}{\micro\watt}$ average power was focused on the sample using a 100x/0.95NA objective resulting in an estimated spot diameter of $2r = 2\lambda/(\text{NA}\cdot\pi)\approx\SI{0.36}{\micro\metre}$. The same objective was used to collect the signal in reflection geometry. \autoref{subfig:rtpl_a} shows the measured PL intensity map (longpass 700\,nm) presented in Fig. 3b of the main text for logarithmic intensity scaling. Further, \autoref{subfig:rtpl_b} shows measured PL intensity spectra from six different positions in the same region of the scan as indicated by circles.
The uniformly distributed bright PL signal originates from the pristine regions of \ML{} as confirmed by the PL intensity spectra measured in positions 1, 2, 4, and 5. The respective spectra show a pronounced peak at 730\,nm wavelength which is a mixed contribution of the neutral (X) and charged exciton (X$^-$) in \ML{}. A large triangular region in the center of the WSe$2$ crystal shows an about one order of magnitude lower PL intensity and can be attributed to 2L-WSe\textsubscript{2} as seen from the spectrally shifted peak in the PL intensity spectra measured in positions 3 and 6. In several smaller triangular regions no PL signal different from the intrinsic emission of the substrate can be observed. As the visual contrast of these regions is also similar as for the bare substrate (see Fig. 3a of the main text), we conclude that these regions are holes in the WSe\textsubscript{2} crystal.\\
Further, a regular square pattern of spots is visible in the confocal scan where each spot coincides with the location of an individual nanoantenna. The nanoantenna regions appear relatively brighter on the substrate region without \ML{} showing a PL enhancement of the autofluorescence from the substrate due to the nanoantennas. In contrast, the PL from \ML{} appears relatively weaker in the position of the nanoantennas which we attribute to the interplay of the resonant nanoantennas with guided modes in the silicon dioxide slab of the substrate. Depending on their geometrical parameters and respective spectral resonance position, the nanoantennas can facilitate an in- or out-coupling from farfield radiation (travelling out of the subtrate plane) to the guided modes in the substrate (travelling in the substrate plane). This interplay leads to a different out-coupling efficiency of PL from emitters depending on their relative position, i.e. in the substrate or on top of the nanoantennas. For comparison,~\autoref{subfig:rtpl_c} and~\autoref{subfig:rtpl_d} show the measured PL intensity map and spectra, respectively, for an array with slightly different geometrical nanoantenna parameters. In this case, the relative brightness of PL at the location of the nanoantennas is enhanced further suggesting that the effect of guided modes in the substrate is sensitive to the exact resonant conditions of the nanoantennas.\\
Interestingly, in all spectral measurements we find an enhanced PL intensity from WSe$_2$ for the positions of the nanoantennas (orange curves) as compared to bare substrate (blue curves) irrespective of the observations in the confocal scans. Note that the confocal scan were performed with a pixel integration time of 5\,ms while the signal for the spectral point measurements was integrated for several 10s of seconds. This might indicate that the PL response of the WSe$_2$ on substrate varies over a time scale of milliseconds to seconds which is much slower than electronic processes (picoseconds to nanoseconds). Such effects are typically related to charge-trapping in localized defect states. As a qualitative statement we therefore conclude that the presence of the nanoantennas can enhance the radiative decay rate of defect states in WSe$_2$. \autoref{subfig:rtpl_e} shows a measured decay curve of PL from \ML{} on bare substrate (blue curve) and on top of the nanoantenna array (orange curve) from the same sample as shown in~\autoref{subfig:rtpl_a} where we have used a 730/10\,nm bandpass filter. We observe a two-component decay and by fitting a two-exponential function we extracted PL lifetimes of $(2.62\pm 0.05)$\,ns and $(0.37\pm 0.01)$\,ns on bare substrate. The decay constants reduce to $(1.90\pm 0.06)$\,ns and $(0.32\pm 0.01)$\,ns on the nanoantenna array which further shows the presence of a nearfield interaction of the nanoantennas with emitters in \ML{}.
\section{Spectral analysis of the excitonic photoluminescence from \ML{}}
We have analyzed the excitonic content of the measured PL spectra from \ML{} at cryogenic conditions ($T=3.8\,K$) using a multi Voigt-line fitting procedure (see for example Bender et al.~\cite{Bender2023spectroscopic}). For 1L-WSe$_2$, both in-plane (bright) and out-of-plane (grey/dark) excitons may contribute to the PL emission owing to the negative conduction band splitting in W-based 1L-TMDs.~\cite{Kosmider2013large,Kormanyos2014spinorbit}
\begin{figure}[h!]
	\centering
	\includegraphics[width=13.5cm]{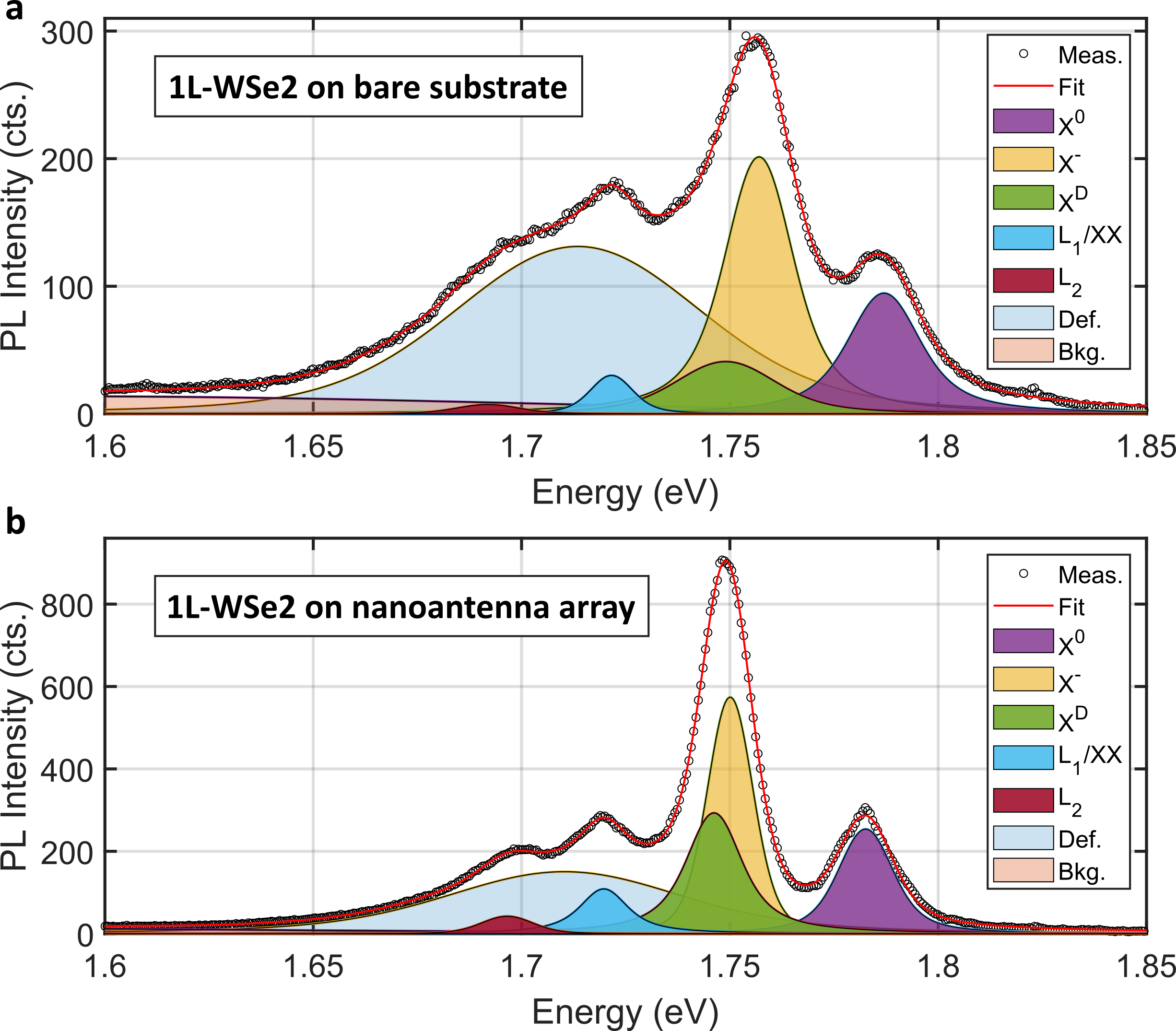}
    {\phantomsubcaption\label{subfig:exciton-spec_a}}
    {\phantomsubcaption\label{subfig:exciton-spec_b}}
	\caption{\textbf{Voigt-profile spectral analysis.} Spectral fitting of the measured cryogenic photoluminescence spectra from 1L-WSe$_2$ on (\subref{subfig:exciton-spec_a}) the bare substrate and (\subref{subfig:exciton-spec_b}) the nanoantenna array using $\sigma^+$ polarized excitation and detection.}
	\label{fig:exciton-spec}
\end{figure}
In our cryo-PL spectral measurements of 1L-WSe$_2$ on the bare substrate (see black circles in~\autoref{subfig:exciton-spec_a} and blue area in Fig. 4b of the main text), we observe several spectral contributions that we attribute as follows:~\cite{You2015observation, Huang2016probing, Robert2017fine, Courtade2017charged, Molas2019probing, Jindal2025brightened} The highest-energy peak corresponds to the neutral bright exciton $X^0$ (purple area) at a wavelength of \qty{693.8}{\nm} (\qty{1.787}{\eV}). The next-highest-energy peak spectrally overlaps with the expected energies of negatively charged trion states, here represented by one contribution $X^-$ (yellow area) at \qty{705.7}{\nm} (\qty{1.757}{\eV}), as well as grey/dark exciton states $D^0$ and trion states $D^-$, again represented by one contribution $X^D$ (green area) at \qty{708.9}{\nm} (\qty{1.749}{\eV}). In the main text, we labeled this peak as $X^-/D$ with its maximum intensity appearing at \qty{706.1}{\nm} (\qty{1.756}{\eV}). On the low-energy side of the spectrum, we observe a broad defect-mediated emission band (light blue area) centered at \qty{723.4}{\nm} (\qty{1.714}{\eV}) and an almost flat background signal (light pink area) with its center wavelength lying outside of the spectral range of interest. We also observe one additional peak that spectrally overlaps with the expected energy of a localized defect state $L_1$ and bi-exciton states ($XX$), here represented by one contribution $L_1/XX$ (cyan area) at \qty{720.0}{\nm} (\qty{1.722}{\eV}). In the main text, we labeled this peak as $L_1/XX$ with its maximum intensity appearing at \qty{720.4}{\nm} (\qty{1.721}{\eV}). We note that the bi-exciton is not expected to be dominant at the relatively low pump fluences used in this work.~\cite{You2015observation,Huang2016probing} We observe another minute feature attributed to a different localized defect state $L_2$ (dark red area) at \qty{732.8}{\nm} (\qty{1.692}{\eV}). The sum of all contributions (red curve) agrees well with the measured spectrum.\\
For 1L-WSe$_2$ on the nanoantenna array (see black circles in~\autoref{subfig:exciton-spec_b} and orange area in Fig. 4b of the main text), we observe a similar spectral composition. However, the $X^0$, $X^-/X^D$, and $L_1/XX$ peaks are shifted toward smaller energies by \qty{4}{\meV}, \qty{7}{\meV}, and \qty{3}{\meV}, respectively. We attribute the smaller spectral shifts of \qtyrange[range-units=single]{3}{4}{\meV} to natural variations across the monolayer area and potential modifications in the local strain due to the non-planarized sample surface. Interestingly, for the $X^-/X^D$ peak we observe a larger spectral shift owing to the relatively enhanced dark exciton contribution $X^D$ for 1L-WSe$_2$ on the nanoantenna array (compare green areas). This enhancement results from coupling to the strong out-of-plane nearfield components mediated by the plasmonic nanoantenna.\\ \\
Further, we note that we observe a similar DOCP of PL for the $X^0$ and $X^-/D$ peaks (compare with Fig. 4c of the main text) on both the bare substrate and the nanoantenna array. As the circular polarization contrast of emission is related to valley-polarized in-plane excitons, we conclude that the $X^0$ and $X^-/D$ peaks exhibit similar in-plane contributions, indicating that the dark exciton contribution $X^D$ is dominated by grey trions.~\cite{Jindal2025brightened}
\section{Comparison of experimentally and numerically obtained angular circular dichroism maps}
\begin{figure}[h!]
	\centering
	\includegraphics[width=\textwidth]{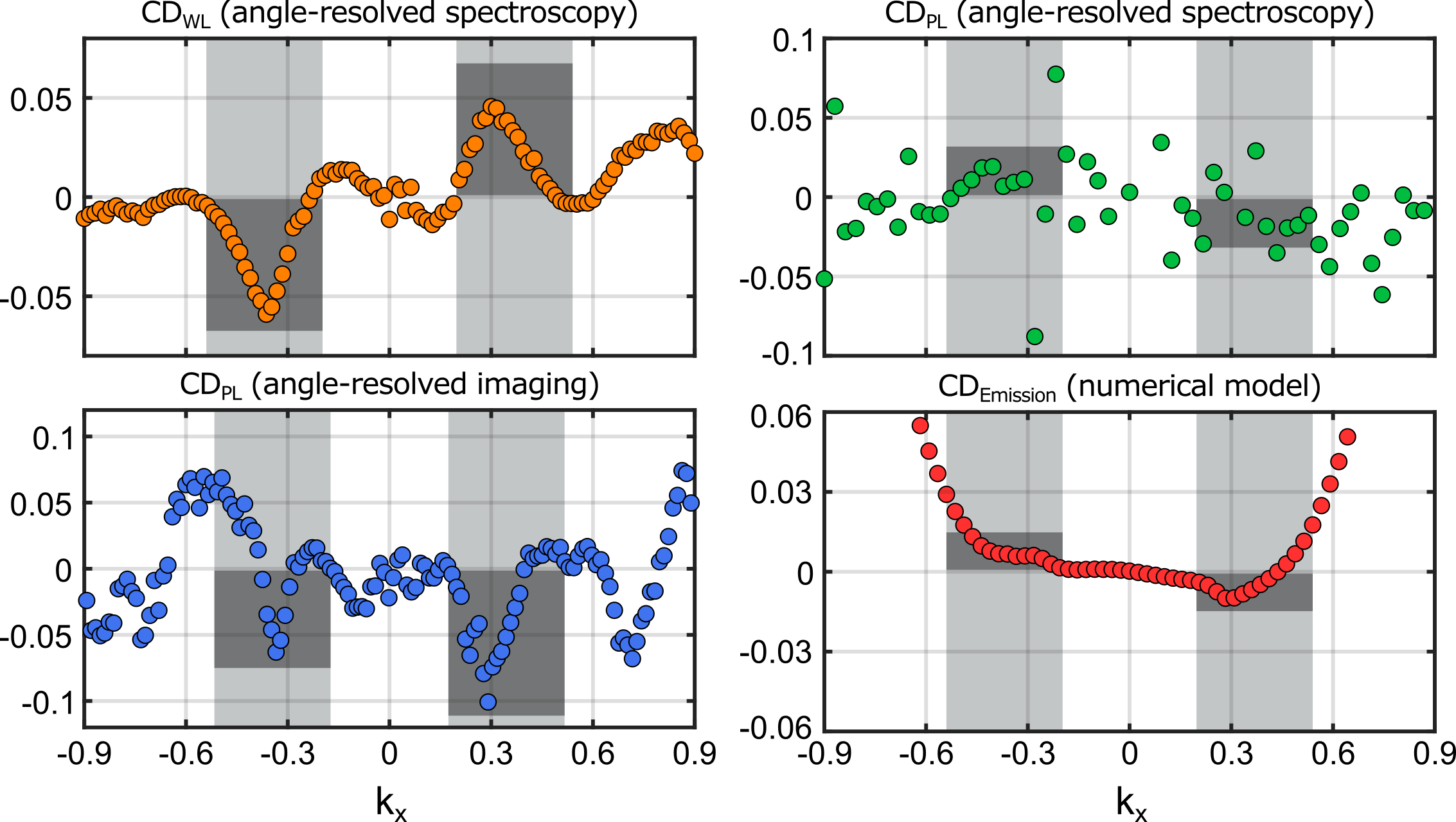}
	\caption{\textbf{Comparison of CD distributions.} Angular circular dichroism of 1L-WSe$_2$ on top of the nanobar dimer array along the $k_x$-direction, obtained at a wavelength of 710 nm by angle-resolved white light (orange markers) and photoluminescence (green markers) spectroscopy, angle-resolved photoluminescence imaging (blue markers), and numerical emission modeling (red markers). The light and dark gray areas were added as a guide to the eye.}
	\label{fig:CD-comparison}
\end{figure}

We compare the angular circular dichroism (CD) of 1L-WSe$_2$ on top of the nanobar dimer array obtained by experimental and numerical methods. \autoref{fig:CD-comparison} shows the angular CD along the $k_x$-direction ($k_y=0$), obtained at a wavelength of 710 nm by angle-resolved white light (orange markers) and PL (green markers) spectroscopy, angle-resolved PL imaging (blue markers), and numerical emission modeling (red markers).
For angle-resolved white light spectroscopy, we find a prominent antisymmetric distribution with respect to the $k_x$-direction appearing at the angular range marked by the light-gray area. For clarity, we further highlighted the amplitudes of the antisymmetric feature by the dark gray areas. As discussed in the main text, for the angle-resolved PL spectroscopy result, the distribution is less systematic. However, a significant antisymmetric feature with reduced amplitudes appears in the same angular range. Interestingly, the sign of the PL feature is reversed with respect to the white light result. Note that we also observe a sign reversal of the angular CD in PL by our numerical emission model when varying the filling factor of the monolayer (see Fig. 6d of the main text). Hence, the orientation of these antisymmetric features likely depends on the specific nearfield interaction for a given emitter distribution.\\
In experiments, the whole unitcell of the array is covered with 1L-WSe$_2$ (100\% filling factor). The numerical emission model for this case confirms the angular range of this antisymmetric feature. However, our model predicts a lower amplitude than observed in the angle-resolved spectroscopy measurements. This may indicate that in experiments a stronger nanoantenna response is observed than numerically predicted for a fully covered unitcell, likely caused by local excitation and emission enhancement limited to the close proximity of the nanoantenna.\\
Ultimately, we compare the angle-resolved PL spectroscopy measurements to our angle-resolved PL imaging results. Note that we limit our discussion to a qualitative comparison, as both data sets were obtained under different experimental condition: for angle-resolved PL imaging, we employed a 710/10 nm bandpass filter and a tightly focused excitation beam (0.46 $\upmu$m spot diameter), while for the angle-resolved PL spectroscopy the spectral resolution is determined by the spectrometer (three-pixel spectral width of 0.7 nm) and we employed a widened excitation beam (5.5 $\upmu$ spot size).\\
The PL imaging result shows larger magnitudes of the CD distribution but no antisymmetric features within the angular range discussed above. We note that pronounced minima appear in the same angular range that relate to diffractive modes. However, their amplitude difference follows the same trend as the antisymmetric features observed for the other PL results.
\section{Numerically calculated angular emission patterns}
\label{sec:emission-model-intensities}
\begin{figure}[h!]
	\centering
	\includegraphics[width=0.75\textwidth]{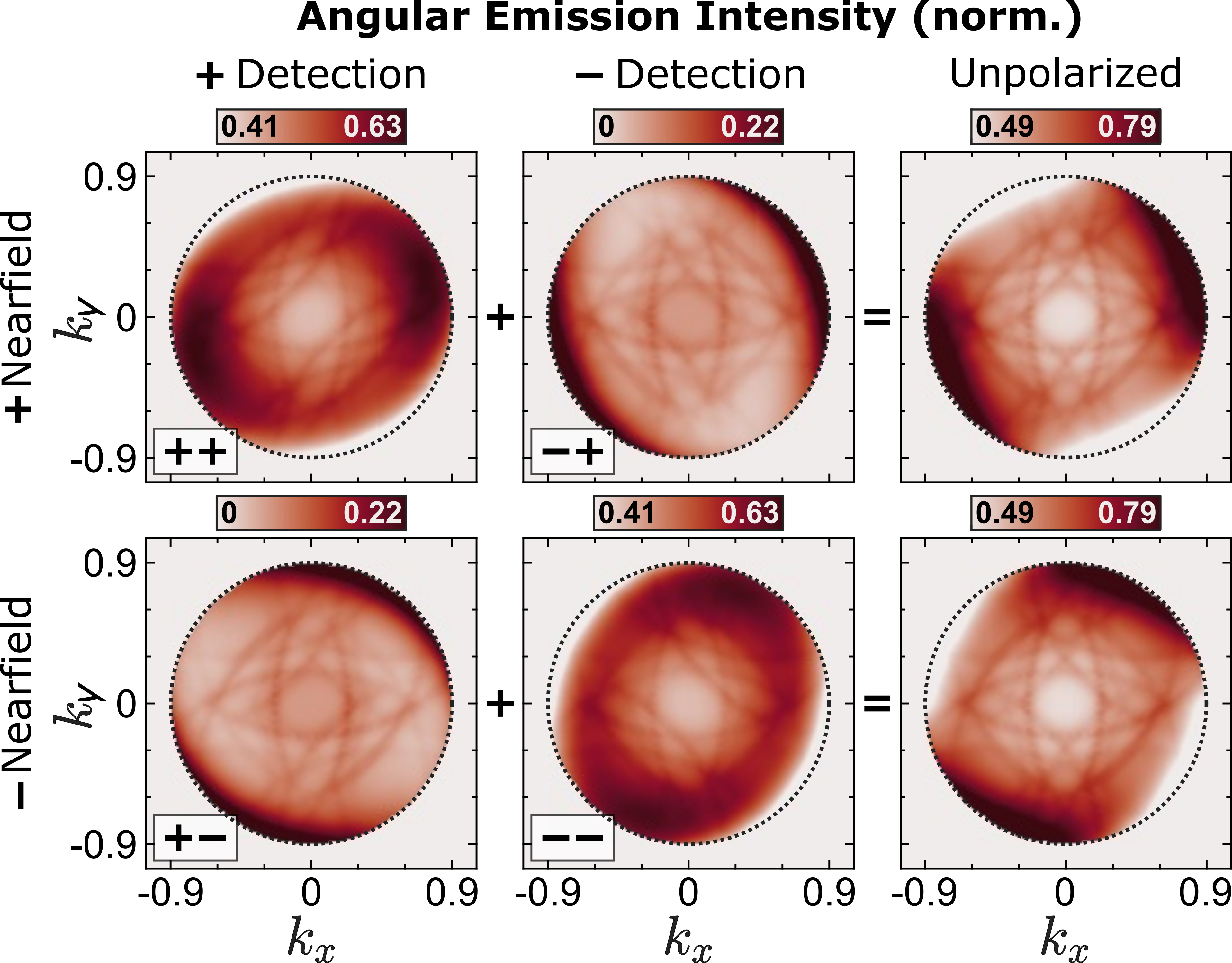}
	\caption{\textbf{Emission modeling.} Numerically calculated angular emission intensities for the nanobar dimer array, obtained at a wavelength of 710 nm, for circular nearfield polarization (rows) and farfield polarization (left and middle columns) polarizations, as well as for unpolarized farfield (right column).}
	\label{fig:emission-model-intensities}
\end{figure}

\autoref{fig:emission-model-intensities} shows the numerically calculated angular emission intensities obtained for the nanobar dimer array, as discussed in the main text. For simplicity, we introduce a Jones notation to distinguish between the circular polarization states as follows: $\sigma^{+}_{\text{farfield}}\,|\,\sigma^{-}_{\text{nearfield}}\equiv{+-}$, with analogous definitions for all other polarization combinations.
The calculated emission patterns are in very good agreement with the experimentally obtained angular PL intensities shown in Fig.~5 of the main text. The modeled emission patterns show a clear influence by diffractive modes arising from the periodic arrangement of the nanoantennas within the array. Depending on the circular polarization in the farfield, these diffractive modes appear with different strengths, giving rise to similar polarization-dependent farfield patterns as experimentally observed and discussed in the main text.
\section{Extrinsic chirality of the nanobar dimer array}
\begin{figure}[h!]
	\centering
	\includegraphics[width=0.7\textwidth]{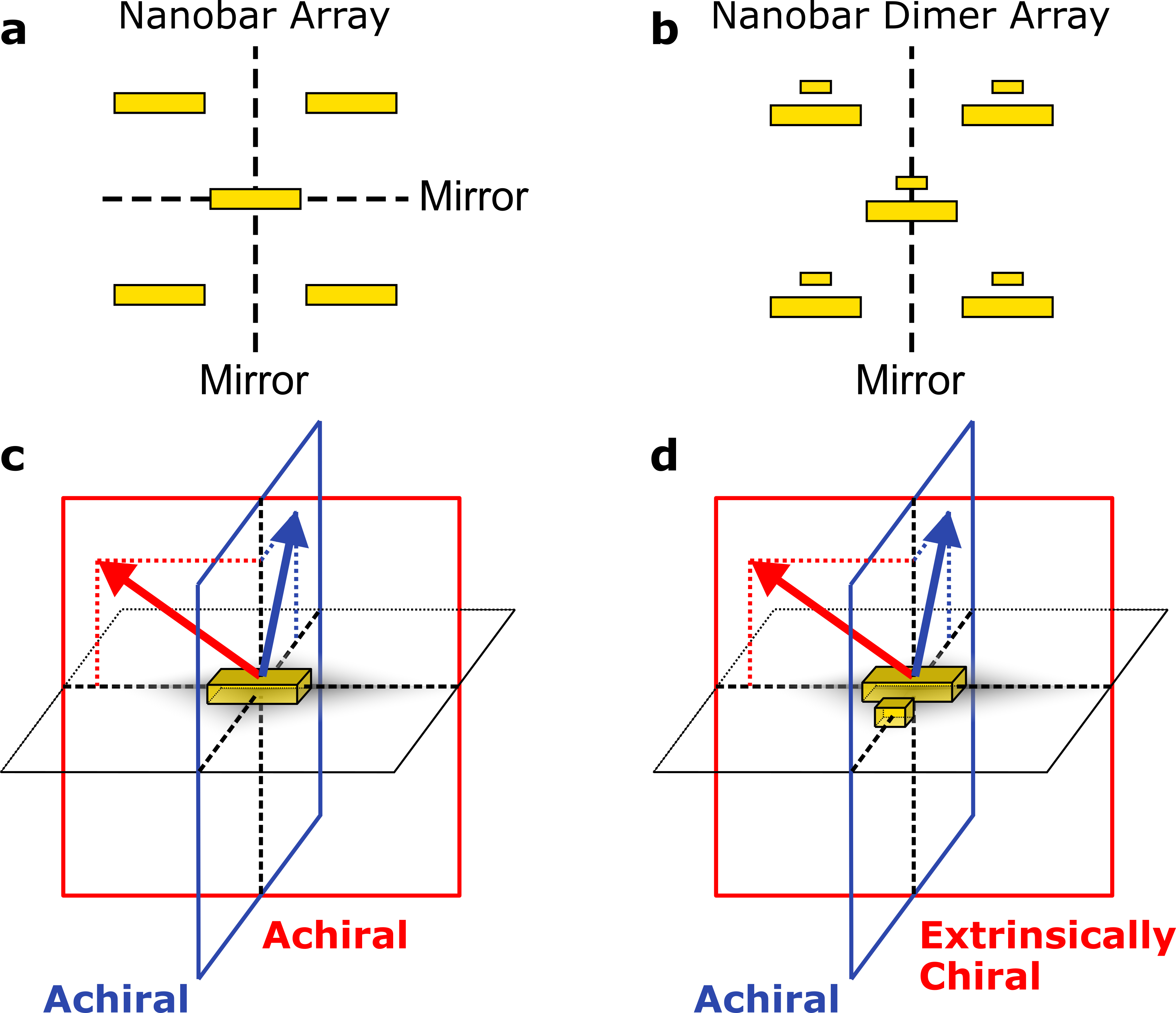}
    {\phantomsubcaption\label{subfig:extrinsic-chirality_a}}
    {\phantomsubcaption\label{subfig:extrinsic-chirality_b}}
    {\phantomsubcaption\label{subfig:extrinsic-chirality_c}}
    {\phantomsubcaption\label{subfig:extrinsic-chirality_d}}
	\caption{\textbf{Symmetry properties and extrinsic chirality.} Sketch of an achiral array of (\subref{subfig:extrinsic-chirality_a}) single nanobars and (\subref{subfig:extrinsic-chirality_b}) nanobar dimers. Symmetry planes and properties for oblique emission angles for the (\subref{subfig:extrinsic-chirality_c}) single nanobar array and the (\subref{subfig:extrinsic-chirality_d}) nanobar dimer array.}
	\label{fig:extrinsic-chirality}
\end{figure}

In the following discussion, we analyze the geometrical and electromagnetic chiral properties of the nanoantenna arrays, as studied in the main text, and within the context of circularly polarized field polarizations.\\
True structural chirality requires a lack of any mirror plane in three dimensions,~\cite{Caloz.2020} and as Caloz et al. emphasize, "chiral exclusively refers to phenomena in materials composed of particles with structural handedness".~\cite{Caloz.2020} In such inherently chiral structures, magnetoelectric coupling (bianisotropy) is the fundamental origin of any optical handedness. \cite{Hentschel.2012l1,Caloz.2020}\\
A single gold nanobar possesses $C_{2}$ symmetry, and a square array of nanobars also retains $C_{2}$ symmetry, whenever the nanobar axis aligns with any lattice axis, as shown in~\autoref{subfig:extrinsic-chirality_a}.
Geometrically, dimers of two non-identical nanobars can retain mirror symmetry (C$_\mathrm{S}$-symmetric) with respect to one axis, as shown in~\autoref{subfig:extrinsic-chirality_b}, rendering the dimer achiral in the strict, three‐dimensional sense.\\
Along the normal direction (and for small angular deviations), the single nanobar and the nanobar dimer exhibit the aforementioned mirror symmetry in the plane, hence their angular CD must vanish for $\mathbf{k}$‐vectors near $k_x=k_y=0$.~\cite{Plum.2016} In contrast, for oblique emission angles the effective symmetry is lowered: each emitted plane wave sees a "tilted" array, and thereby breaking the mirror symmetry with respect to its "plane of incidence".\\
As discussed above, the single nanobar array and the nanobar dimer array posses different symmetry axes. Hence, we analyze the effect of oblique angles with respect to these axes, as shown in~\autoref{subfig:extrinsic-chirality_c} and~\autoref{subfig:extrinsic-chirality_d}, respectively. As highlighted by the red and blue arrows, any oblique $\mathbf{k}$‐vector that lies within one of the symmetry planes (red and blue), is breaking the mirror symmetry with respect to the plane normal to its plane of incidence. For the single nanobar array, however, mirror symmetry is always retained due to the second mirror axis, thus rendering the system achiral. In contrast, for the nanobar dimer, an oblique $\mathbf{k}$‐vector that lies within the plane normal to its mirror plane (red arrow), breaks the two‐dimensional mirror symmetry in three‐dimensional space. This is the hallmark of \emph{extrinsic} chirality.~\cite{Plum.2016}\\
Moreover, the observed chiral asymmetry is not generated by the dimer alone but by the entire layered structure. For circularly or elliptically polarized light, the silicon substrate acts as polarization‐rotating mirror. Due to the thin spacer it interacts in the near field with the nanobar layer.~\cite{Sperrhake.2021} Furthermore, Plum et al. have shown that, additionally, for planar, achiral resonators on top of a reflector, oblique incidence can produce enhanced chiral effects.~\cite{Plum.2016} In our system, the angular CD signal is most robust when the plane of incidence aligns with the dimer axis, directly linking the handedness of illumination ($\sigma^+$ versus $\sigma^-$) to directional emission via reciprocity.
\section{Angle-integrated farfield intensity contrast}
The angular emission intensities show a rich behavior when considering different emission angles as well as for different monolayer filling factors. Integrating intensities over many emission angles, hence may significantly lower the intensity contrast. However, as discussed in the main text and in Sec.~S.7 of the Supporting Information, the angular asymmetry is robust with respect to the two halfspaces that are linked to extrinsic chirality. Thus, integrating all intensity contributions within each of these halfspaces, should retain a finite intensity contrast.
\begin{figure}[h!]
	\centering
	\includegraphics[width=0.6\textwidth]{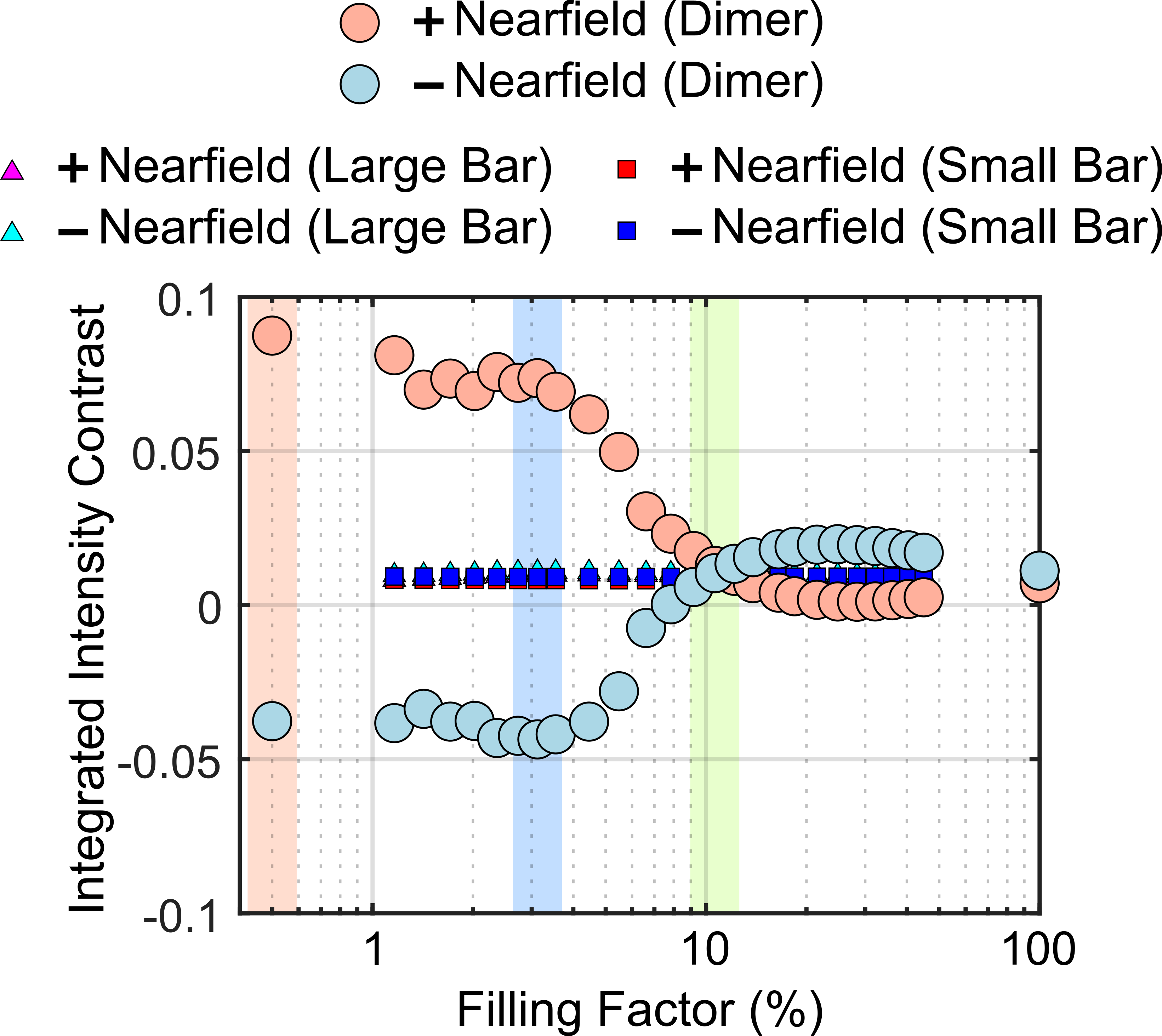}
	\caption{\textbf{Angle-integrated farfield intensity contrast.} Numerically calculated half-space intensity contrast, obtained from $\sigma^+$ and $\sigma^-$ polarized nearfield components, for the nanobar dimer array (red and blue circles), the large-nanobar array (magenta and cyan triangles), and the small-nanobar array (red and blue squares).}
	\label{fig:integrated-intensity-contrast}
\end{figure}
In addition to the fixed-direction intensity contrast provided in Fig. 6 of the main text, we calculated the contrast between these integrated intensities, as shown in~\autoref{fig:integrated-intensity-contrast}. The resulting curve clearly shows a finite contrast for the nanobar dimer array. Additionally, the distribution is similar to the result obtained for a fixed farfield direction along the $k_x$ axis (compare Fig.~6d of the main text), as a consequence of the symmetry-protected extrinsic chirality of the two halfspaces.\\
Importantly, the single nanobar arrays show no significant angle-integrated intensity contrast. This confirms the initial assumption that angle-integration may drastically reduce the observed intensity contrast in the absence of extrinsic chirality.
\section{Farfield circular polarization contrast}
\begin{figure}[h!]
	\centering
	\includegraphics[width=0.6\textwidth]{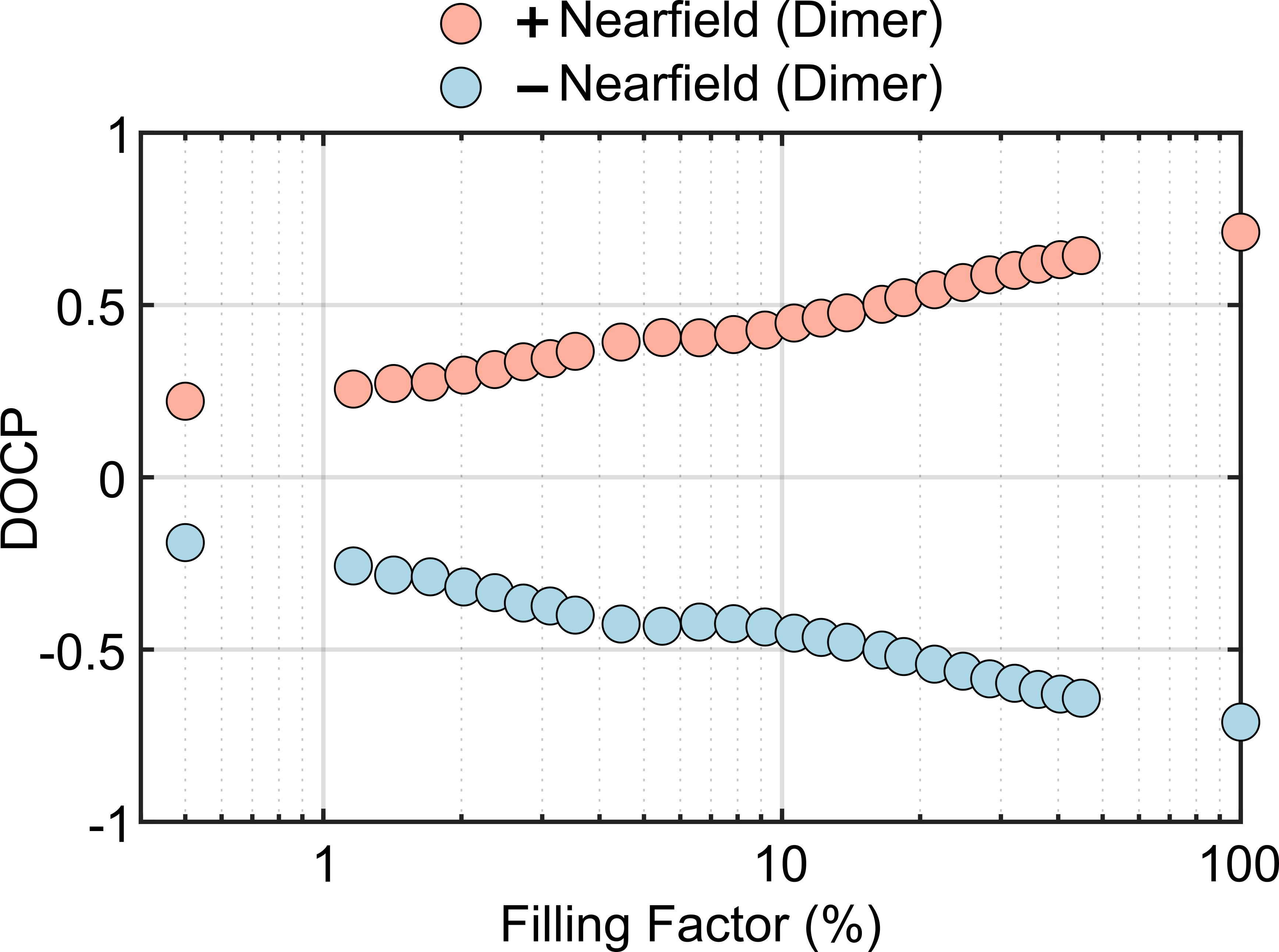}
	\caption{\textbf{Farfield circular polarization contrast.} Numerically calculated degree of circular polarization of the nanobar dimer array, obtained from the angle-integrated emission patterns for $\sigma^+$ and $\sigma^-$ polarized nearfield components.}
	\label{fig:farfield-circular-contrast}
\end{figure}

Besides valley-selective directional effects, we have also analyzed the farfield circular polarization of the light emitted from the nanoantenna array. For this, we consider the calculated angular emission intensities (e.g. compare~\autoref{fig:emission-model-intensities}) and obtain their overall radiated powers $P_{++}$, $P_{-+}$, $P_{+-}$, and $P_{--}$ by integration over all emission angles within the numerical aperture. \autoref{fig:farfield-circular-contrast} shows the respective DOCP in the farfield, defined as $\mathrm{DOCP}_\pm = (P_{+\pm} - P_{-\pm})/(P_{+\pm} + P_{-\pm})$, obtained for a given nearfield polarization $\pm$ according to the Jones notation introduced in~\autoref{sec:emission-model-intensities}.
We find that the DOCP gradually decreases with decreasing filling factor, showing that the nanoantenna does not preserve the circular polarization upon farfield scattering. For larger filling factors, the nanoantenna's relative contribution with respect to emission from the whole unitcell, however, becomes less significant, leading to the observed increase in the DOCP. Note that the DOCP without the nanoantenna, or for an infinitely large unitcell, has a magnitude close to $1$.\\ \\
In conclusion, we find that valley-selective emitters being located within the nearfield region of the nanoantenna, and therefore most strongly interacting with the nanoantenna, exhibit the lowest circular polarization contrast in the farfield. Hence, the valley-information of the localized emitters is obscured in farfield polarization measurements. Importantly, our nanoantenna design allows discerning the valley-information from directional contrast measurements, whose magnitude, in contrast, is larger for smaller filling factors.
%
%
\newpage
\bibliography{references}